\documentclass[aps,amsmath,amssymb,superscriptaddress,nofootinbib,preprintnumbers,twocolumn,showpacs]{revtex4-1}
\usepackage{epsfig}
\usepackage{bbm}
\usepackage{color}
\usepackage{mathrsfs,graphicx,bm,amsmath}

\newcommand{\Tr}{\mathrm{Tr}}

\newcommand{\bq}{\textbf{q}}

\begin{document}

\title{Quantum Field Theory for the Three-Body Constrained Lattice Bose Gas \\
\ Part II: Application to the Many-Body Problem}
\author{S. Diehl}
\affiliation{Institute for Quantum Optics and Quantum Information of the Austrian Academy of Sciences, A-6020 Innsbruck, Austria}
\affiliation{Institute for Theoretical Physics, University of Innsbruck, A-6020 Innsbruck, Austria}
\author{M. A. Baranov}
\affiliation{Institute for Quantum Optics and Quantum Information of the Austrian Academy of Sciences, A-6020 Innsbruck, Austria}
\affiliation{Institute for Theoretical Physics, University of Innsbruck, A-6020 Innsbruck, Austria}
\affiliation{RRC ``Kurchatov Institute'', Kurchatov Square 1, 123182 Moscow, Russia}
\author{A. J. Daley}
\affiliation{Institute for Quantum Optics and Quantum Information of the Austrian Academy of Sciences, A-6020 Innsbruck, Austria}
\affiliation{Institute for Theoretical Physics, University of Innsbruck, A-6020 Innsbruck, Austria}
\author{P. Zoller}
\affiliation{Institute for Quantum Optics and Quantum Information of the Austrian Academy of Sciences, A-6020 Innsbruck, Austria}
\affiliation{Institute for Theoretical Physics, University of Innsbruck, A-6020 Innsbruck, Austria}

\begin{abstract}
We analyze the ground state phase diagram of attractive lattice bosons, which are stabilized by a three-body onsite hardcore constraint. A salient feature of this model is an Ising type transition from a conventional atomic superfluid to a dimer superfluid with vanishing atomic condensate. The study builds on an exact mapping of the constrained model to a theory of coupled bosons with polynomial interactions, proposed in a related paper \cite{Diehl09I}. In this framework, we focus by analytical means on aspects of the phase diagram which are intimately connected to interactions, and are thus not accessible in a mean field plus spin wave approach. First, we determine shifts in the mean field phase border, which are most pronounced in the low density regime. Second, the investigation of the strong coupling limit reveals the existence of a ``continuous supersolid'', which emerges as a consequence of enhanced symmetries in this regime. We discuss its experimental signatures. Third, we show that the Ising type phase transition, driven first order via the competition of long wavelength modes at generic fillings, terminates into a true Ising quantum critical point in the vicinity of half filling.
\end{abstract}

\pacs{03.75.Hh,03.75.Kk,11.15.Me,67.85.Hj,64.70.Tg}

\maketitle


\section{Introduction}

It was recently recognised that two-body and three-body loss processes for bosons in an optical lattice could give rise to effective models involving two-body and three-body hardcore constraints, respectively. The two-body case was observed in an experiment with Feshbach molecules \cite{Syassen08, Ripoll09}, while it has been proposed theoretically to take advantage of strong three-body loss to create a three-body hardcore constraint in bosonic  \cite{Daley09,Roncaglia09} and fermionic \cite{HanKantian09} lattice systems. The mechanism behind the constraint is that the dissipative process suppresses coherent tunnelling processes that would create double or triple occupation and lead to loss. 

A salient feature of a bosonic lattice gas with three-body onsite constraint is the possibility to tune it to attractive two-body interactions. The associated dimer bound state formation has a profound effect on the many-body system, resulting in an Ising-type quantum phase transition from a conventional atomic superfluid to a dimer superfluid with vanishing atomic order parameter but nonzero pairing correlation. The possibility of observing Ising type behavior in cold atomic gases has been uncovered earlier by Radzihovsky \emph{et al.} \cite{Radzihovsky04,Radzihovsky07} and Romans  \emph{et al.} \cite{Sachdev04} in the context of resonant Bose gases in the continuum, i.e. at low densities. This, however, turns out to be challenging due to the poor stability of the molecular Bose gas close to the resonance \cite{Naegerl09}. Here, we encounter a weak coupling analog of this scenario on the lattice, in which the stabilization of the system is provided by the blockade mechanism leading to the 3-body hardcore constraint. Besides this feature, the presence of the lattice leads to intriguing enrichments compared to the continuum physics, as we will demonstrate in this paper. 

The qualitative picture for the Ising transition can be obtained within a simple Gutzwiller approach, in which the three-body constraint is easily built in via choice of the ansatz wave function \cite{Daley09}. However, this treatment leaves a number of questions unanswered, which arise on various length scales in the problem, and misses out important -- even qualitative -- aspects of the phase diagram as we will show. On the microscopic scale, this concerns the bound state formation, as well as the correct form of the  effective theory for dimers in the strong coupling limit. On the intermediate scales, relevant to the thermodynamics, one may wonder to what extent the phase border obtained within the mean field is quantitatively accurate. Finally, a thorough analysis of the competition of the long range low energy degrees of freedom is necessary to answer the question of the true nature of the phase transition. We note that all these effects are tied to interactions, thus not available in a simple spin wave extension of the mean field theory.
\begin{figure}
\begin{center}
\includegraphics[width=0.9\columnwidth]{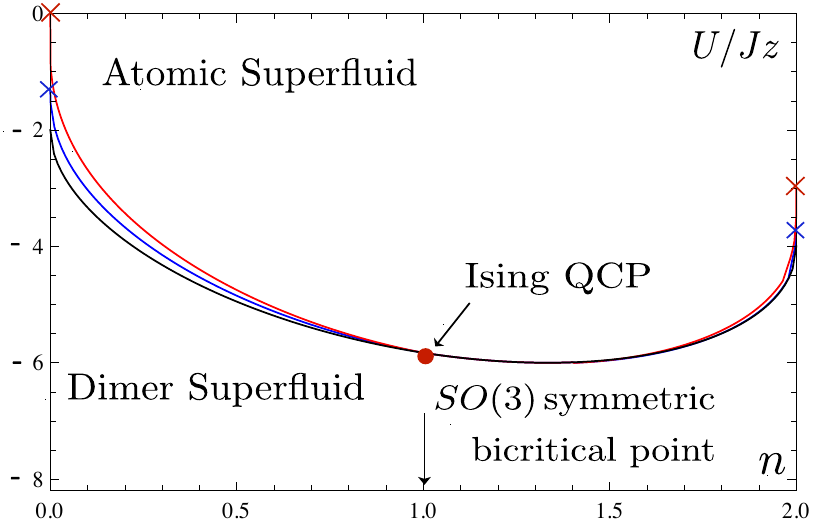}
\end{center}
\caption{\label{PhaseDiagram} Phase diagram for the attractive 3-body hardcore constrained attractive Bose-Hubbard model. The black curve below is the mean field result, while red and blue curve correspond to $d=2,3$ where fluctuations are included. The bound state formation of dimers ($n=0$) and di-holes ($n=2$) takes place at the red and blue crosses for $d=2,3$, determining the endpoints of the critical lines (Sec. \ref{sec:ASFDSF}). A bicritical point, characterized by energetically degenerate but different orders (superfluid and charge density wave) is reached asymptotically at half filling. It can be detected experimentally ramping a superlattice (Sec. \ref{sec:DSFCDW}). An Ising quantum critical point, connecting the two ordered phases, is predicted in the vicinity of half filling, while the correlation length is large but finite away from this point (Sec. \ref{sec:IR}).}
\end{figure}

This paper is the second one of a sequence of two related papers. In Ref. \cite{Diehl09I}, we have developed a quantum field theoretical framework which makes it possible to analytically address the above questions in two and three spatial dimensions. It is based on an exact mapping of the constrained lattice boson model to a coupled theory of two unconstrained bosonic degrees with polynomial interactions. In the related paper \cite{Diehl09I}, we have concentrated on the formal development of this mapping, and performed calculations in the ``vacuum limits'' corresponding to zero and maximum filling $n=0,2$, which are characterized by the absence of spontaneous symmetry breaking. In the present paper we apply this formalism to the many-body problem. We concentrate on the three interaction-related aspects of the many-body problem mentioned above: First, we address the quantitative question of shifts in the phase border, with the result that they are pronounced at low densities, while basically absent as the filling increases to its maximum $n=2$. Second, making use of the perturbative results obtained in \cite{Diehl09I}, we consider the many-body physics in the strong coupling regime, and predict the existence of a new collective mode at half filling $n=1$, whose presence results from a symmetry enhancement from the conventional phase rotation symmetry $U(1)\simeq SO(2)$ exhibited by bosonic systems to an $SO(3)$ symmetry. We propose an experiment to test this scenario, exploring its consequences both analytically as well as using exact numerical methods in one dimension. These studies lead us to call the system in this regime a ``continuous supersolid'' -- a supersolid with a tunable ratio between the superfluid and the charge density wave order parameters (cf. an analogous phenomenon in magnetic \cite{Kohno97,Batrouni99} and attractive fermion \cite{Zhang90} systems). Third, in a long wavelength analysis the phase transition turns out to be first order for generic fillings due to the Coleman-Weinberg mechanism \cite{Coleman73}. This is in line with the low density continuum analysis, which has been carried out in detail in \cite{Radzihovsky07}. In our constrained lattice system, however, we find that the radiatively induced first order transition terminates into a true Ising quantum critical point in the vicinity of half filling, which connects the two ordered phases of atomic and dimer superfluid. Its origin may be traced back to a zero crossing of the dimer compressibility together with a sequence of Ward identities, thus being protected by symmetry. An estimate of the correlation length suggests a broad domain of intermediate fillings $1/2\lesssim n \lesssim 3/2$ on which the correlation length greatly exceeds the dimensions of typical optical lattices, suggesting that the Ising quantum critical behavior could be experimentally observed. Our analytical approach enables us to elucidate the mechanisms behind all our findings, establishing that the latter two effects are unique features of the three-body constraint. Our main results are summarized in the phase diagram presented in Fig. \ref{PhaseDiagram}.

The paper is organized as follows. In Sec. 
\ref{sec:QFT} we first review the steps that lead from the constrained theory to the interacting boson theory. We then  prove Goldstone's theorem for the effective action obeying the constraint principle, and formulate the equation of state. In Sec. \ref{sec:ASFDSF}, we pass on to the calculation of the phase border beyond mean field. Sec. \ref{sec:DSFCDW} discusses the many-body physics in the strong coupling limit, and in Sec. \ref{sec:IR} we investigate the nature of the phase transition by performing the long wavelength limit of the effective action. Our conclusions are drawn in Sec. \ref{sec:Conclusion}.

A summary of our results together with a closer discussion of experimental realizations is presented in \cite{Diehl09Short}.

\section{Quantum Field Theory for the Many-Body Problem}
\label{sec:QFT}

In this section we address two aspects which are particularly relevant for the many-body physics and have not been discussed in \cite{Diehl09I}: The realization of Goldstone's theorem in our constrained model, and the equation of state. To prepare for this discussion and set the notation, we review the construction of the quantum field theory in Sec. \ref{sec:Review}, also making the paper rather self-contained. The reader familiar with the construction, and the reader who is more interested in the physics results of this work, may jump this section. 

\subsection{Review of the Construction}
\label{sec:Review}

The starting point for our analysis is the Bose-Hubbard model with a
three-body onsite hardcore constraint,
\begin{eqnarray}
H\hspace{-0.1cm}=\hspace{-0.1cm}-J\sum_{\langle i,j\rangle}a_{i}^{\dagger
}a_{j}\hspace{-0.1cm}-\hspace{-0.1cm}\mu\sum_{i}\hspace{-0.1cm}\hat{n}%
_{i}+\tfrac{1}{2}U\sum_{i}\hspace{-0.1cm}\hat{n}_{i}(\hat{n}_{i}%
-1),\;a^{\dag\,3}\hspace{-0.05cm}\equiv\hspace{-0.05cm}0,
\end{eqnarray}
Here, $a_i,a_i^\dag$ are the bosonic creation and annihilation operators, $J$ is the hopping matrix element $\mu$ the chemical potential and $U$ the onsite interaction energy. The summation in the first term is performed over nearest neighbors.
Because of the constraint, the original bosonic onsite Hilbert space is
reduced to the three states $|\alpha\rangle,\alpha=0,1,2$.

Following Altman and Auerbach \cite{Altman02}, we introduce three operators
which generate the three onsite states,
\begin{eqnarray}
|\alpha\rangle=t_{\alpha,i}^{\dag}|\text{vac}\rangle=(\alpha!)^{-1/2}\,\,\left(  a^{\dag
}\right)  ^{\alpha}|\text{vac}\rangle,\,\,\sum_{\alpha}t_{\alpha,i}^{\dag}t_{\alpha,i}=\mathbf{1}%
\end{eqnarray}
from some auxiliary ``vacuum" state $|$vac$\rangle$. The operators are not
independent but obey a holonomic constraint as indicated above. The
Hamiltonian in terms of operators $t_{\alpha}$ reads
\begin{align}
H  &  =-J\sum_{\langle i,j\rangle}\big[K_{i}^{(10)}K_{j}^{(10)\dag}%
+2K_{i}^{(21)\dag}K_{j}^{(21)}\label{HtComp}\\
&  +\sqrt{2}(K_{i}^{(21)}K_{j}^{(10)\dag}+K_{i}^{(10)}K_{j}^{(21)\dag
})\Big]\nonumber\\
&  -\mu\sum_{i}\left(  \hat{n}_{1,i}+2\hat{n}_{2,i}\right)  +U\sum_{i}\hat
{n}_{2,i},\nonumber
\end{align}
where
\[
K_{i}^{(10)}=t_{1,i}^{\dag}t_{0,i},\quad K_{i}^{(21)}=t_{2,i}^{\dag}%
t_{1,i},\quad\hat{n}_{\alpha,i}=t_{\alpha,i}^{\dag}t_{\alpha,i}.
\]
Note that in this representation of the constrained Hamiltonian, the
conventional roles of interaction and hopping are reversed: while the
interaction enters the quadratic part of the Hamiltonian, the hopping term
gives rise to effective kinematic interactions. The representation is
therefore ideally suited in a strong coupling limit.

In a naive Gross-Pitaevski treatment of the Hamiltonian, achieved by replacing
the operators with complex valued amplitudes $t_{\alpha,i}\rightarrow
f_{\alpha,i}$ in Eq. \eqref{HtComp}, reproduces precisely the Gutzwiller
mean-field energy, i.e. a classical Hamiltonian field theory for spatially
varying amplitudes $f_{\alpha,i}$, where the holonomic constraint $\sum
_{i}f_{\alpha,i}^*f_{\alpha,i}=1$ is the normalization of the onsite wave function. We
now show how one can introduce a convenient description of the theory on the
quantum level. To illustrate the method we consider the case of vanishing
density\textbf{\ }$n\rightarrow0$ (the generalization to an arbitrary density
$0\leq n\leq2$ will be given below). In this limit, it is convenient to express
the $t_{0,i}$ operators in terms of $t_{1,i}$ and $t_{2,i}$ operators using the
constraint. Writing $t_{0,i}=|t_{0,i}|\exp\mathrm{i}\varphi_{i}$, we
observe that the phase $\varphi_{i}$ is unphysical: it can be eliminated via a
\emph{local} redefinition of the remaining operators, $t_{\alpha,i}%
=t_{\alpha,i}\mathrm{e}^{\mathrm{i}\varphi_{i}}(\alpha=1,2)$. Thus, we may
consider $t_{0}$ as real, and replace $t_{0,i}=X_{i}^{1/2}$, $X_{i}=1-\hat
{n}_{1,i}-\hat{n}_{2,i}$ in $K_{i}^{(10)}$. Obviously, the square roots are
impracticable for any quantum field theory because they give rise to vertices of
arbitrarily high order. To eliminate this problem, we use the fact that the
matrix elements of $X_{i}^{1/2}$ and $X_{i}$ on our subspace are the same:
either $1$ or $0$. Consequently, on the subspace we may replace
\begin{eqnarray}
K_{i}^{(10)}=t_{1,i}^{\dag}\sqrt{X_{i}}\rightarrow t_{1,i}^{\dag}X_{i},\quad
X_{i}=(1-\hat{n}_{1,i}-\hat{n}_{2,i}),
\end{eqnarray}
and analogous for the hermitian conjugate. More formally, the replacement can
be justified by noting that the constraint operator is a projection, $X_{i}^{2}=X_{i}$, and that the Taylor representation for a function of
such an operator is $f(X)=f(0)(1-X)+Xf(1)$ \footnote{See \cite{Diehl09I} for a subtlety in deriving this formula.}. With this implementation of the
constraint, the remaining operators $t_{1},t_{2}$ can be treated as
\emph{standard bosonic operators} acting in a complete Hilbert space
$\mathcal{H}=\prod_{i}\mathcal{H}_{i}$, where $\mathcal{H}_{i}=\{|n_{i}%
\rangle|m_{i}\rangle\},n_{i},m_{i}=0,1,...$ is a bosonic Hilbert space for
\textquotedblleft atoms\textquotedblright\ $t_{1}$ and \textquotedblleft
dimers\textquotedblright\ $t_{2}$ at each site $i$ : $|n_{i}\rangle=\left(
n_{i}!\right)  ^{-1/2}(t_{1}^{\dag})^{n_{i}}|0\rangle_{i}$ and $|m_{i}%
\rangle=\left(  m_{i}!\right)  ^{-1/2}(t_{2}^{\dag})^{m_{i}}|0\rangle_{i}$.
The onsite Hilbert spaces $\mathcal{H}_i$ can naturally be splitted into a physical
subspace $\mathcal{P}_i$ with $n_{i}+m_{i}=0$ or $1$, and an orthogonal
unphysical one $\mathcal{U}_i$ with $n_{i}+m_{i}>1$, $\mathcal{H}_i=\mathcal{P}_i%
\oplus\mathcal{U}_i$. Important for our construction is that the Hamiltonian $H$
has no matrix elements between physical and unphysical subspaces, $\langle
u|H|p\rangle=\langle p|H|u\rangle=0$, where $|p\rangle\in\mathcal{P}=\prod_{i}\mathcal{P}_{i}$ and
$|u\rangle\in\mathcal{U}=\prod_{i}\mathcal{U}_{i}$, and, therefore, is block diagonal,
$H=H_{\mathcal{P}}+H_{\mathcal{U}}$. As a result, these subspaces do not mix
during evolution, and all quantities, both dynamical and statistical,
factorize. For example, for the partition function one has
\begin{align}
Z  &  =\mathrm{Tr}\exp(-\beta H)=Z_{\mathcal{P}} + Z_{\mathcal{U}}\\
&  =\sum\limits_{\{p\}}\langle p|\exp(-\beta H_{\mathcal{P}})|p\rangle + 
\sum\limits_{\{u\}}\langle u|\exp(-\beta H_{\mathcal{U}})|u\rangle.\nonumber
\end{align}
Consequently, if we find a way to discriminate between the physical and unphysical contributions, we may indeed conceive the operators $t_{1,2}$ as conventional bosonic ones. 

Such a setting is provided by using the \emph{effective action} to encode the physical information of the theory, see e.g. \cite{AmitBook}. It is defined as the Legendre transform of the free energy $W[j] = \log Z[j]$ (we introduce a source term $j= (j_1,j_1^\dag, j_2, j_2^\dag)$ and use $\hat\xi = (t_1^\dag ,t_1, t_2^\dag , t_2)$):
\begin{eqnarray}
\Gamma [\xi] = - W[j] + \int j^T \xi, \quad \xi \equiv \frac{\delta W[j] }{\delta j} ,
\end{eqnarray}
where the new variable $\xi = \langle \hat \xi\rangle$ is the field expectation value or the ``classical'' field. The effective action has the following representation in terms of a functional integral,
\begin{eqnarray}\label{EAPI}
\exp - \Gamma[\xi] =\hspace{-0.2cm} \int \hspace{-0.15cm}\mathcal D \delta \xi \exp - S [ \xi + \delta \xi] \hspace{-0.1cm}+ \hspace{-0.15cm}\int \hspace{-0.15cm}j^T \delta \xi ,  j = \frac{\delta \Gamma [\xi]}{\delta\xi }.
\end{eqnarray}
where $\delta \xi  \equiv \hat \xi  - \xi$, and the Euclidean action $S = \int_\tau\sum_i t_1^\dag \partial_\tau t_1 + t_2^\dag \partial_\tau t_2 + H[t_1,t_2]$. The Hamiltonian now is to be interpreted as a function for classical though fluctuating, time dependent fields. The last identity in Eq. \eqref{EAPI} is the full quantum equation of motion, and the equilibrium situation we are interested in is specified by $j=0$, where no mixing between the physical and the unphysical sector occurs. Usually, the most general form of the effective action is only restricted by the symmetries of the microscopic theory. Since, as shown above, no couplings mapping from $\mathcal U \leftrightarrow \mathcal P$ are generated, we have identified a means to distinguish physical vs. unphysical contributions by writing down the most general form for the effective action for the physical sector by directly excluding couplings which would violate this constraint. 

Now we generalize the procedure to arbitrary density. We first follow \cite{Altman07} but then apply our exact procedure. While we have so far replaced the $t_0$ operator, which generates the mean field vacuum state $|\Omega\rangle = \prod_i t_{0,i}^\dag |\text{vac}\rangle$, we now consider a more general mean field vacuum,\begin{eqnarray}\label{amplitudesI}
|\Omega \rangle &=& \prod_i \big(\sum\limits_\alpha r_\alpha \exp ( i \alpha\phi) |\alpha\rangle_i\big) \\\nonumber
&=& \prod_i \big(\sum\limits_\alpha r_\alpha \exp ( i\alpha\phi) t_{\alpha,i}^\dag \big)|\text{vac}\rangle 
\stackrel{!}{=} \prod_i b_{0,i}^\dag  |\text{vac}\rangle.
\end{eqnarray} 
For site independent amplitude moduli $r_\alpha$, these states allow for the description of homogeneous ground states with spontaneous phase symmetry breaking: If, e.g., all $r_\alpha \neq 0$, the requirement of a fixed spontaneously chosen overall condensate phase $\phi$ requires the phase relation $\theta_\alpha =\alpha \phi$; this fixed phase relation is the manifestation of spontaneous symmetry breaking in the Fock space. We can now introduce a new set of operators $b^\dag_\alpha$ ($\alpha =0,1,2$) in which $b_0^\dag$ creates the mean field vacuum and will be eliminated. Such a transformation is performed via a two-parameter unitary rotation, whose rotation angles are chosen such that the new operators fluctuate around the new vacuum state and do not feature expectation values,
\begin{eqnarray}\label{ratation}
b_{\alpha,i}^\dag = (R_\theta R_\chi)_{\alpha\beta} t_{\beta,i}^\dag
\end{eqnarray}
with the explicit form of the rotation matrices
\begin{eqnarray}\label{RotMat}
R_\theta &=&  \left(
\begin{array}{ccc}
  {\cos \theta/2} & {0} & {\sin \theta/2\mathrm e^{2\mathrm i \phi}}\\
  {0} & {1} & {0}\\
  {-\sin \theta/2 \mathrm e^{-2\mathrm i \phi}} & {0} & {\cos \theta/2}
\end{array}
\right), \\\nonumber
 R_\chi   &=&  \left(
\begin{array}{ccc}
  {1} & {0} & {0}\\
  {0} & {\cos \chi/2} & {-\sin \chi/2\mathrm e^{\mathrm i \phi}}\\
  {0} & {\sin \chi/2 \mathrm e^{-\mathrm i \phi}} & {\cos \chi/2}
\end{array}
\right) .
\end{eqnarray}
A finite $\theta (\chi)$ corresponds to a finite amplitude in $|2\rangle (|1\rangle)$ and we will see below how these quantities are fixed via the Goldstone theorem. The precise relation is
\begin{eqnarray}\label{amplitudesII}
r_0 = \cos \theta/2, \,\, r_1 = \sin \theta/2 \sin \chi/2, \,\, r_2 =  \sin \theta/2 \cos \chi/2.\nonumber\\
\end{eqnarray}
At this point we can repeat the steps described above for the case $n=0$ in complete analogy. The constraint is implemented via the replacement
\begin{eqnarray}\label{REP}
 b_{0,i}  \to X_i \equiv 1-  b_{1,i}^\dag  b_{1,i} -  b_{2,i}^\dag  b_{2,i},\quad 
b_{0,i}^\dag b_{0,i} \to X_i .
\end{eqnarray}
The second line is simply a rearrangement of the holonomic constraint. The resulting bosonic Hamiltonian, which is then quantized by means of a functional integral, is rather complex, and we will analyze it below. However, it exhibits a simple structure,
\begin{eqnarray}\label{HDecomp}
H = E_{\text{GW}} +  H_{\text{SW}} + H_{\text{int}}.
\end{eqnarray}  
$E_{\text{GW}}$ is the Gutzwiller mean field energy and $H_{\text{SW}}$ describes the quadratic spin wave theory \footnote{A linear contribution, as naively expected in the expansion about the condensate, does not occur due to Goldstone's theorem, see below.}. The corrections to the mean field phase diagram, as well as nontrivial effects in the deep infrared physics which we are interested in here, are not captured at this quadratic level. They are all encoded in the interaction part $H_{\text{int}}$.

Choosing the qualitative form of the ground state prerequisites a certain knowledge about the physics of the system. Equipped with the right qualitative ground state, we can then perform quantitative calculations beyond the mean field level based on our mapping. Indeed, Eq. \eqref{HDecomp} suggests an interpretation of our construction as an exact requantization procedure of the Gutzwiller mean field theory. This is in complete analogy to the conventional treatment of e.g. bosonic continuum systems with broken symmetries, where in a first step a certain order parameter is chosen and the theory is expanded around it. However, on the lattice the right choice of the qualitative features of the ground state might be less obvious. For example, spatial modulations of the order parameter are possible, such as exhibited by charge density waves. This is easily incorporated in the formalism, and such a situation will be indeed encountered in Sec. \ref{sec:DSFCDW}.

In sum, we have obtained the following simple result: supplying the most general form of the effective action with a constraint principle, the evaluation can proceed as in a standard polynomial boson theory. Similar to symmetries, the restrictions on the full theory leverage over from the microscopic theory. Unlike symmetries, the relevance of the constraint depends on scale, being restrictive on short distances, while on long distances power counting arguments lead to an effectively unconstrained though interacting spin wave theory with two degrees of freedom (see below). In practical computations, we can evaluate a theory of standard coupled bosonic fields. This opens up the powerful toolbox of modern quantum field theoretical methods for calculations in onsite constrained models. 

\subsection{Goldstone's Theorem and the Constraint Principle}
\label{sec:GSConstraint}

We will derive Goldstone's theorem from a comparison of the full quantum equation of motion and the effective potential. The latter is defined as the homogeneous part of the effective action, obtained by inserting temporally and spatially homogeneous field configurations $\mathcal U =  \Gamma[b_1^\dag, b_1, b_2^\dag,b_2]/V_{d+1}$ ($V_{d+1}=M^d/T$ is the quantization volume, $M$ the number of lattice sites in each lattice direction). The possible dependences of the effective action and potential on the fields are strongly restricted by both symmetry and constraint principle; the effective potential is further limited by the requirement of homogeneity. We will show here that the constraint leads to an additional $U(1)$ invariant on which the effective potential may depend with no counterpart in unconstrained theories, but we will also demonstrate that it does not break the validity of Goldstone's theorem -- in line with the intuition that the microscopic constraint would not affect the long wavelength physics too strongly. Note, however, that the constraint has an impact on the long wavelength physics, as it is indirectly responsible for the presence of the Ising quantum critical point close to unit filling $n=1$ (cf. Sect. \ref{sec:SymmArg}). Therefore, a thorough discussion of Goldstone's theorem seem adequate.

Let us construct the most general dependence of the effective potential on the variables $b_1,b_2$. For simplicity of the presentation, we focus on a spontaneously broken symmetry for the dimers ($\theta\neq 0,\pi$), while the atoms are in the normal phase ($\chi =\ 0,\pi$). The latter field can therefore be excluded from the following considerations. There are two possible terms associated to the original $t_2$ degree of freedom that might appear in the effective action: Either it appears as a local combination $\hat n_{2,i} = t_{2,i}^\dag t_{2,i}$, or as a bilocal (in general, $n$-local) combination, such that the constraint has to be taken into account via proper combination with $t_0$, e.g. $t_{2,i}^\dag t_{0,i}$. While the local combination respects the $U(1)$ symmetry, the second term must appear with a conjugate partner as $t_{2,i}^\dag t_{0,i} t_{0,j}^\dag t_{2,j}$. In order to implement the finite density, we now apply the rotation prescription $t_{2,i} = s b_{0,i} + c b_{2,i}, t_{0,i} = c b_{0,i} - s^* b_{2,i}$ and subsequently impose the constraint $b_{0,i} \to X_i$. (Here and in the following, we abbreviate $s = \sin \theta/2 \mathrm e^{-2\mathrm i \phi}, c= \cos \theta/2$.) Now we specialize to the homogeneous part of the effective action: We Fourier transform the operators and restrict to the zero frequency and momentum part of the combinations. We then find that the effective potential can be written as a function of two invariants, 
\begin{eqnarray} \label{UandInvars}
&&\mathcal U ( \rho, \lambda^\dag \lambda ),\\\nonumber
 \rho &=& (s X +  c b_2^\dag ) (s X  + c b_2 ) \\\nonumber
 &=& s^2 + cs (Xb_2 + b_2^\dag X ) + (c^2-s^2)b_2^\dag b_2 - s^2 b_1^\dag b_1, \\\nonumber
 \lambda &=& (s X +  c b_2^\dag )(c X -  s b_2 )  \\\nonumber
 &=& cs  + c^2 b_2^\dag X - s^2 Xb_2  - 2 cs b_2^\dag b_2 - cs b_1^\dag b_1,
\end{eqnarray}
where $X,b_2$ denote the zero momentum and frequency components of these field expressions, and without loss of generality we have chosen $s$ real. Note that neglecting the constraint by setting $X \to 1$, and considering low densities, $s\to \theta/2, c\to 1$, we recover the standard quadratic form for the condensate density from the local combination, $\rho = \hat b_2^\dag \hat b_2$, with $\hat b_2 = s  + b_{2}$. However, the constraint principle requires a more complicated form of $\rho$, as well as the account for a second invariant $\lambda^\dag \lambda$. In the following, we will be concerned with first and second derivatives of the effective potential with respect to $b_2,b_2^\dag$, which are evaluated at the physical point $b_2=b_2^\dag = b_1=b_1^\dag=0$. Thus, we may set $X\to 1, b_1,b_1^\dag \to 0$ from the outset. Now we will show that the most general dependence of $\mathcal U$ can be further restricted.  For $b_{2,i}(\tau), b^\dag_{2,i}(\tau)$, we introduce the basis of hermitian fields $\sigma_{i}(\tau),\pi_{i}(\tau)$,
\begin{eqnarray}\label{HermField}
b_{2,i} (\tau) &=& \frac{1}{\sqrt 2}\,\, \Big(\sigma_{i}(\tau)  + \mathrm i \pi_{i}(\tau) \Big), \\\nonumber
b_{2} (q) &=& \frac{1}{\sqrt 2}\,\, \Big(\sigma(q)  + \mathrm i \pi(-q) \Big), 
\end{eqnarray}
which as the original fields do not carry expectation values. Here we have used the Fourier conventions 
\begin{eqnarray}
b_{2, i}(\tau) &=& \int_q e^{\mathrm i q x_i} b_{2}(q) , \quad b^\dag_{2, i}(\tau) = \int_q e^{-\mathrm i q x_i} b^\dag (q), \\\nonumber
 x_i &=& (\tau, \textbf{x}_i),\quad q = (\omega,\textbf{q}), \quad \int_q = \int\frac{d\omega}{2\pi} \sum_\textbf{q}.
\end{eqnarray}
We calculate the local combination in terms of these operators, 
\begin{eqnarray}
\rho = s^2  + (c^2 - s^2) \tfrac{1}{2} (\sigma^2 + \pi^2) + \sqrt{2} cs \sigma,
\end{eqnarray}
($\sigma = \sigma(q=0), \pi = \pi (q=0)$) and we observe that $\lambda^\dag \lambda$, to the relevant quadratic order, can be written as 
\begin{eqnarray}
\lambda^\dag \lambda  = s^4  + (c^2 - s^2) \rho - 2(cs)^2 \sigma^2.
\end{eqnarray}
Thus, the most general dependence of the effective potential on the homogeneous fields $\sigma,\pi$ is given by
\begin{eqnarray} 
\mathcal U ( \rho, \sigma^2).
\end{eqnarray}

Now we study the mass matrix, which can be calculated from the effective potential as the second derivative with respect to $\sigma,\pi$. In particular, the form of the effective potential implies for the $\pi$   mass or gap
\begin{eqnarray}\label{BIG}
\frac{\partial^2\mathcal U(\rho,\sigma^2) }{\partial \pi\partial \pi }\Big|_{\sigma= \pi =0} = \Big(\frac{\partial^2 \rho }{\partial \pi^2} \mathcal U'  + \Big(\frac{\partial \rho }{\partial \pi}\Big)^2 \mathcal U'' \Big)\Big|_{\sigma= \pi =0} 
\end{eqnarray}
(primes denote derivatives w.r.t. the invariant $\rho$) with
\begin{eqnarray}\label{derivs}
\frac{\partial \rho }{\partial \pi} = 0 , \quad 
\frac{\partial^2 \rho }{\partial \pi^2} = 
 c^2 -  s^2.
\end{eqnarray}
To complete the derivation of Goldstone's theorem, we calculate the equation of motion for $\sigma$ from the effective action, but immediately specialize to the case of homogeneous fields, 
\begin{eqnarray}\label{EoMSigma}
\frac{\delta \Gamma}{\delta \sigma_{i}(\tau)}\Big|_{\text{hom}} = \frac{\partial \mathcal U(\rho,\sigma^2)}{\partial \sigma} = 2\sigma \frac{\partial \mathcal U(\rho,\sigma^2)}{\partial \sigma^2}  + \sqrt{2} c s\,\,  \mathcal U' \stackrel{!}{=} 0.\nonumber\\
\end{eqnarray}
By construction we have $\sigma=0$ as the solution of the equation of motion, and furthermore in the presence of spontaneous symmetry breaking $c s\neq 0$. Thus 
\begin{eqnarray}\label{GSM}
\mathcal U' =0.
\end{eqnarray}
This simple relation indicates the presence of the gapless Goldstone mode: The $\pi$ gap calculated in Eq. \eqref{BIG} vanishes due to Eq. \eqref{GSM}, and since $\partial \rho /\partial \pi = 0$, cf. Eq. \eqref{derivs}. This property is protected by the $U(1)$ symmetry of the problem. Though the form of the effective potential is more complicated than in the continuum at low densities, where the effective potential depends only on the low density limit of the invariant $\rho$, we can explicitly prove Goldstone's theorem.

Note, that the equation of motion \eqref{EoMSigma} for $\sigma$ also excludes any homogeneous linear term in this field. The same is true for the $\pi$ mode. Such terms would not be compatible with the equilibrium condition \eqref{EoMSigma}. This excludes nonzero couplings from the homogeneous terms $\sim \sigma, \pi$ or $b_2,b_2^\dag$ in the effective action. Furthermore, via Eq. \eqref{UandInvars} the linear terms $b_2,b_2^\dag$ are connected by the constraint principle to cubic terms: only the combinations $\sim b_2 X , X b_2^\dag$ occur in the effective potential. Thus, combining Goldstone's theorem and the constraint principle, we see that the cofficients of the terms  $ b_2 X , X b_2^\dag$ must vanish, i.e.: 
\begin{eqnarray}\label{HomVan}
&&\frac{\delta^3 \Gamma}{\delta b_{2,i}^\dag \delta b_{2,i} \delta b_{2,i}}\Big|_{\text{hom}} = \frac{\delta \Gamma}{\delta b_{1,i}^\dag \delta b_{1,i} \delta b_{2,i}}\Big|_{\text{hom}} \\\nonumber
&=& \frac{\delta \Gamma}{\delta b_{2,i}^\dag \delta b_{2,i} \delta b_{2,i}^\dag}\Big|_{\text{hom}} = \frac{\delta \Gamma}{\delta b_{1,i}^\dag \delta b_{1,i} \delta b_{2,i}^\dag}\Big|_{\text{hom}} = 0.
\end{eqnarray}

In the presence of an atomic condensate $\chi \neq 0$, analogous equilibrium conditions can be derived for the $b_1$ field. 

In the symmetric phases $n=0,2 (\theta =0,\pi)$, no distinction between the phase and the amplitude mode appears and the mass matrix is degenerate. In this case, Goldstone's theorem reduces to the condition for the existence of a dimer/di-hole bound state.

\subsection{Equation of State}
\label{sec:EoS}

The equation of state is obtained as the average over the particle number operator $\hat N = \sum_i \hat n_i, \hat n_i = \hat n_{1,i} + 2 \hat n_{2,i}$. Thus, after rotation we have for the particle density   
\begin{eqnarray}\label{FullDens}
 n =\langle \hat N\rangle/M^d &=& 2 |s|^2 + (c^2- |s|^2)[ \langle b_1^\dag b_1\rangle + 2 \langle b_2^\dag b_2\rangle] \nonumber\\
 &&+ 2 c [s^* \langle X b_2\rangle + s \langle b_2^\dag X \rangle] = - \frac{\partial \mathcal U}{\partial \mu} .
\end{eqnarray}
The second equality results from the path integral representation of the effective action, and is due to the coupling $-\mu \hat N$ in the microscopic action. The connected two-point functions are given by the traces of the full Green's functions for $b_1$ and $b_2$. A convenient shorthand to relate the connected Green's functions to its one-particle irreducible counterpart is $\xi = ( b_1^\dag,b_1,b_2^\dag,b_2)$,
\begin{eqnarray}\label{ConnDens}
\langle b_1^\dag b_1 \rangle &=& \langle \xi_1 \xi_2 \rangle =  \Tr \,\, G_{12} , \quad \langle b_2^\dag b_2 \rangle = \langle \xi_3 \xi_4 \rangle =  \Tr \,\, G_{34} ,\nonumber\\
G_{ab} &=& (\Gamma^{(2) \,-1})_{ab}, \quad \Gamma^{(2) }_{ab} = \frac{\delta^2 \Gamma}{\delta \xi_a \delta \xi_b } .
\end{eqnarray}
where we suppress spatial or momentum indices. Tr runs over these as well as over the internal (field space) indices. More explicit formulae will be discussed in the next section. Furthermore, the three-point correlation is related to the one-particle irreducible three-point vertex via Eq. \eqref{FullDens}  (cf. e.g. \cite{AmitBook})
\begin{eqnarray}
\langle X b_2\rangle &=& - \langle b_1^\dag b_1 b_2\rangle - \langle b_2^\dag b_2 b_2\rangle  \\\nonumber
&=& \sum_{a,b,c}\mathrm{Tr}  [  G_{1a} G_{2b}  + G_{3a} G_{4b} ] G_{4c}  \Gamma^{(3)}_{abc},\\\nonumber
\Gamma^{(3)}_{abc} &=&  \frac{\delta^3 \Gamma}{\delta \xi_a \delta \xi_b\delta \xi_c } .
\end{eqnarray}

At this point, we stress that the parameter $|s|^2$ in the equation of state \eqref{FullDens} must not be interpreted as the condensate fraction, though the formal appearance naively suggests such an interpretation. Instead, $2|s|^2$ should be seen as the classical or mean field contribution to the total particle density, and the rest of the equation is due to fluctuations on top of this mean field state. A standard interpretation of the above equation is only possible in the low density limits $n\to 0,2$. Omitting the three-point correlations, Eq. \eqref{FullDens} reduces to leading order to the familiar form from thermodynamics in the continuum for $\theta \to 0$, while taking a similar structure for $\theta \to \pi$, 
\begin{eqnarray}\label{FullDensSimpLow1}
 \theta \approx 0 &:& n = 2 (\delta\theta/2)^2 +  \langle b_1^\dag b_1\rangle + 2 \langle b_2^\dag b_2\rangle, \\\nonumber
 \theta \approx \pi &:& n = 2 - [2 (\delta\theta/2)^2 +  \langle b_1^\dag b_1\rangle + 2 \langle b_2^\dag b_2\rangle ].
\end{eqnarray}
In these cases, $\delta \theta/2$ may be interpreted as the condensate order parameter. We furthermore observe from Eq. (\ref{FullDens}) that around $\theta =\pi/2$ there is a point where fluctuations are strongly suppressed compared to the mean field contribution due to a cancellation. For a proper definition of the condensate fraction in the system, we can use the Gutzwiller expression for the original boson operator expectation, $\langle b_i\rangle = s c$, however with the value of $\theta$ determined from the implicit condition Eq. (\ref{GSM}). More generally, we emphasize that Eqs. (\ref{GSM},\ref{FullDens}) provide the two exact, but implicit conditions that determine the two parameters $\theta,\mu$. A further nonzero expectation value for the the single atom degree of freedom, described by $\chi\neq 0$, adds a further such condition analogous to Eq. \eqref{GSM}. 

Our effective action formalism is capable to describe the system at any finite temperature. Calculations in this regime are beyond the scope of this paper, but let us sketch how the high temperature disordered phase is described within our theory. Increasing the temperature in the system will populate the connected parts of Eq. \eqref{FullDens} increasingly such that at the phase transition to the symmetric phase without symmetry breaking $f_1,f_2\to 0$. In other words, the condensate angles vanish, $\theta, \chi\to 0$. We may interpret this scenario as  the complete population of the ``vacuum amplitude'' $f_0$, which is needed to fulfill the holonomic constraint but does not enter the equation of state. The effect of destruction of the order parameters can also be seen from the condition $m_{\pi}^2 = \partial^2\mathcal U /\partial \pi^2= 0$. A finite temperature will act to generate a positive thermal mass or gap contribution, such that at some temperature there exists no finite $\theta, \chi$ and a gapless mode ceases to exist. At this point, where Goldstones's theorem can no longer be satisfied, the symmetry broken phases become unstable and the system enters the disordered high temperature phase. 

\section{ASF--DSF Phase Border} 
\label{sec:ASFDSF}

In this section we embark the calculation of the phase border. We will study the phase border by approaching it from the dimer superfluid side where there is no atomic condensate, and calculate at which interaction strength the atoms become unstable towards an atomic superfluid. Thus, we first provide the explicit form of the Hamiltonian in the presence of a dimer superfluid, but for atoms in the normal phase. We then consider the low density limits $n\to 0,2$. In these limits, we can establish a controlled small density expansion describing the deviation from exactly $n=0,2$. The central objects for the discussion are the atomic and dimer (di-hole) Green functions, which we know exactly in the limits $n=0,2$ \cite{Diehl09I}. The analysis reveals the intuitive result that the leading many-body effect is a modification of the vacuum ($n=0,2$) Green functions due to the condensate mean field. The dominant fluctuations in these limits are thus vacuum fluctuations renormalizing the Green functions, while the many-body effects can be captured in terms of a Bogoliubov or spin wave theory.  More specifically, we find that vacuum fluctuations strongly modify the relation $\mu(U)$ compared to the mean field relation $\mu(U) =-U/2$, while the role of many-body effects consists mainly in depletion effects in the equation of state. We find that the high energy vacuum fluctuations have a much more pronounced quantitative effect on the phase border than the condensate depletion in the limit $n\to 0$. For $n\to 2$ instead, both effects are rather small, which may be understood in terms of an already tightly bound di-hole state in the region of atom criticality. Based on these insights, we do not expect a strong shift in the phase border in the region $n =1$, which takes place at even stronger coupling, and thus more deeply bound two-particle states. We therefore propose an extrapolation of the scheme from the controlled limits $n\approx 0,2$ to the intermediate regime $n\approx 1$.

\subsection{Rotated Hamiltonian for the dimer superfluid phase} 

Let us now focus on the phase border to the dimer superfluid state. As anticipated above, we address it from the DSF side where $|1\rangle$ is not macroscopically populated and thus $\chi =0$. The kinetic and potential energy operators read, in the new field coordinates,
\begin{eqnarray}\label{RotRels}
K^{(10)}_i\hspace{-0.2cm} &=& b_{1,i}^\dag  (c  b_{0,i} -s^* b_{2,i}) ,\,\, K^{(21)}_i = (s^* b_{0,i}^\dag + c b_{2,i}^\dag ) b_{1,i},\nonumber\\
P^{(11)}_i\hspace{-0.2cm} &=& b_{1,i}^\dag  b_{1,i},\quad P^{(22)}_i = (s^* b_{0,i}^\dag + c b_{2,i}^\dag )(s b_{0,i} + c b_{2,i} ),\nonumber\\
\end{eqnarray}
with $c\equiv \cos \theta/2, s \equiv \sin \theta/2 \mathrm e^{2\mathrm i \phi}$ as above. We can now write the Hamiltonian operator in terms of the new variables, and implement the constraint via $b_{0,i} \to X_i = (1- \hat n_{1,i} -  \hat n_{2,i})$, absorbing the phase of $b_0$ into the remaining two degrees of freedom as discussed in Sec. \ref{sec:Review}. Further making use of the projective property $X_i^2 = X_i$ we find
\begin{widetext}
\begin{eqnarray}\label{HtNew}
H[b_1,b_2] &=& H^{(10)}_{\text{kin}} + H^{(21)}_{\text{kin}} + H^{(\text{split})}_{\text{kin}} +  H_{\text{pot}} ,\\\nonumber
H^{(10)}_{\text{kin}} + H^{(21)}_{\text{kin}} \hspace{-0.2cm}&=&\hspace{-0.1cm} - J \sum_{\langle i,j\rangle}\hspace{-0.1cm}\Big[ (c^2 + 2|s|^2) b_{1,i}^\dag X_i X_j b_{1,j} +(2 |s|^2 + c^2)b_{1,i}^\dag b_{2,i}b_{2,j}^\dag b_{1,j} - 3 c \,\, \big(s b_{1,i}^\dag X_i b_{2,j}^\dag b_{1,j}   +s^* b_{1,i}^\dag b_{2,i}X_j   b_{1,j} \big)    \Big] ,\\\nonumber
H^{(\text{split})}_{\text{kin}} &=& - \sqrt{2} J \sum_{\langle i,j\rangle}\Big[ (c^2 - |s|^2)  b_{2,i}^\dag b_{1,i} X_j b_{1,j}     
 + c \,\, \big(s^*X_i  b_{1,i} X_j b_{1,j}   - s b_{2,i}^\dag b_{1,i}b_{2,j}^\dag b_{1,j}  \big) + \mathrm{h.c.}   \Big], \\\nonumber
H_{\text{pot}} &=&(-2\mu +U)M |s|^2 + (-2\mu +U)(c^2-|s|^2)\sum_i  b_{2,i}^\dag  b_{2,i}  +  [- \mu -(-2\mu +U)|s|^2)]  \sum_i b_{1,i}^\dag  b_{1,i} \\\nonumber
&&+ (-2\mu +U) c \sum_i \Big[s b_{2,i}^\dag X_i + s^*X_i b_{2,i}\Big].
\end{eqnarray}
\end{widetext}
We remind the reader that, as shown in Ref. \cite{Diehl09I} and briefly discussed in Sect. \ref{sec:Review}, these operators $b_{1,2}$ may be interpreted as standard bosonic operators. The cubic term in the second line of $H_{\text{pot}}$ can be omitted from the outset, and we will do this in the following: As argued above, the coefficient of the linear part has to vanish due to the equation of motion, i.e. the equilibrium condition, and the cubic parts are connected to the linear ones via Eq. \eqref{HomVan}, such that their coefficient has to vanish as well. However, this does not exclude the possibility of \emph{nonlocal} cubic terms as they appear in the kinetic terms of the Hamiltonian. The total Euclidean action in the presence of condensation reads
\begin{eqnarray}
S[b_1, b_2] = \int d \tau \Big( \sum_i  b_{1,i}^\dag \partial_\tau  b_{1,i} +   b_{2,i}^\dag \partial_\tau  b_{2,i} + H[ b_1, b_2]\Big),\nonumber\\
\end{eqnarray}
where the Hamiltonian is to be interpreted in the Heisenberg picture and as a function of classical field variables. 
Quantizing this theory with the functional integral leads precisely to the representation of the effective action Eq. \eqref{EAPI}. 


\subsection{Low density limits $n\approx 0,2$} 

In the following, we will analyze the theory in the vicinity of the physical vacua where $n\approx 0,2$, described by $\theta_0 \approx 0, \pi$. The limits $n=0,2$ have been discussed in \cite{Diehl09I} in detail. The Hamiltonians governing these situations describe the scattering of few particles in the absence of many-body effects and can be written as 
\begin{eqnarray}\label{eq:Hlow}
H_n &=&-\sum_{\langle i,j\rangle}\big[g_{n,1} b_{1,i}^{\dag}X_{i}X_{j}b_{1,j} + g_{n,2}  b_{2,j}^{\dag}b_{1,j}b_{1,i}^{\dag}b_{2,i}\nonumber\\
&& +\sqrt{2}J(b_{2,i}^{\dag}b_{1,i}X_{j}t_{1,j}+ b_{1,j}^{\dag}X_{j}b_{1,i}^{\dag
}b_{2,i})\Big]\label{HCubic}\nonumber\\
&&  +\sum_{i}[(U-2\mu_n)\hat n_{2,i}-\mu_n \hat n_{1,i}].
\end{eqnarray}
The operators $b_{1,2}$ represent the bosonic single and two-particle excitations, corresponding to atoms and dimers resp. holes and di-holes. Here, for $n=0$ we have $g_{0,1}  = J,g_{0,2}  = 2J,\mu_0 = \mu$ and for $n=2$, $g_{2,1}  = 2J,g_{2,2}  = J,\mu_2 =- \mu+U$. At these points the exact solution of the (two-body) scattering problem, and thereby an exact calculation of the atomic and dimer Green's function, is available as shown in \cite{Diehl09I}. While the case of the atomic Green function is trivial as there are no renormalization effects in the vacuum, for the dimers/diholes we find the results
\begin{widetext}
\begin{eqnarray}
\label{FullGreen}
G_d^{-1} (\omega;\mu_0,\textbf{k}) &=& U + \Big[ \int\frac{d^dq}{(2\pi)^d} \frac{1}{- 2 (\epsilon_\bq  + \epsilon_{\bq - \textbf{k}}) + \mathrm i \omega - 2\mu_0 } \Big]^{-1},\nonumber\\
G_h^{-1} (\omega;\mu_2,\textbf{k}) &=& U  + \frac{3}{4} (\mathrm i \omega - 2 \mu_2) + \frac{1}{4}\Big[ \int\frac{d^dq}{(2\pi)^d} \frac{1}{- 4 (\epsilon_\bq  + \epsilon_{\bq - \textbf{k}}) +\mathrm i \omega - 2\mu_2 } \Big]^{-1}.
\end{eqnarray}
\end{widetext}
We will now perform a controlled expansion in the condensate angle deviation $\delta \theta\ll 1$ from the special points  $\theta_0 = 0, \pi$. It corresponds to a Bogoliubov approximation for the condensation physics, but with coupling constants obtained from the exact solution of the "vacuum" scattering problem. The procedure amounts to a resummation of ladder diagrams. These vacuum fluctuations are responsible for strong shifts in the phase border as we will see. 

Our expansion is defined with Hamiltonians of the form
\begin{eqnarray}
H &=& H_n +  \frac{\delta\theta}{2} \delta H_n + \mathcal O (\delta \theta^2).
\end{eqnarray}
The additional Hamiltonian $\delta H_n$ generates new scattering vertices which are $\mathcal O (\delta \theta)$. Diagrams with more than one of the new vertices may thus be discarded, and we may restrict our attention to diagrams with at most one of them. They will be discussed in a moment. 

In the low density cases $n \approx 0,2$ the discussion can be lead in parallel, due to the similar mathematical structure of the Hamiltonians. Around $\theta_0 =0$, we replace $c = 1, s= \delta \theta/2$ and the additional Hamiltonian reads 
\begin{eqnarray}\label{HtAddDim}
\delta H_0 &=&- J  \sum_{\langle i,j\rangle}    \sqrt{2} \big[  X_i  b_{1,i} X_j b_{1,j}   -  b_{2,i}^\dag b_{1,i}b_{2,j}^\dag b_{1,j}   \big] \nonumber\\
&& \qquad + X_i  b_{1,i} b_{1,j}^\dag  b_{2,j}   + \mathrm{h. c.}
\end{eqnarray}
Similarly, around  $\theta_0 =\pi$, setting $s = 1, c= -\delta \theta/2$ we find
\begin{eqnarray}\label{HtAddHole}
\delta H_2&=& - \delta H_0.
\end{eqnarray}
Note, that the zero order Hamiltonians $H_n$  are related by a more complicated transformation of parameters, reflecting the absence of a particle-hole symmetry.

Let us now discuss the impact of the additional Hamiltonian. The stability of the ASF phase is encoded in the full atomic mass matrix, i.e. the inverse Green's function $G_1^{-1}(\omega; \mu , \textbf{q})$  at zero frequency and momentum: If the eigenvalues of the mass matrix are all positive, the phase without atomic condensate (i.e. the condensed dimer phase) is stable. The instability towards a state with atomic condensate can thus been inferred from the vanishing of an eigenvalue of this matrix, or
\begin{eqnarray}\label{characteristic}
\det G_1^{-1}(\omega = 0; \mu, \textbf{q}=0)=0.
\end{eqnarray}
Thus we discuss the beyond mean field effects modifying the inverse atomic Green's function. From the exact solutions of the vacuum problems at $\theta_0=0,\pi$ we know that in these limits the inverse atom Green's function is not directly renormalized: there are no diagrams in the vacuum limits which cause renormalization, but clearly, the function $\mu(U)$ entering the atom propagator changes when taking the exact dimer or di-hole Green's function into account. We now concentrate on the effects of $\delta H_n$. At linear order in $\delta \theta$, we find a direct condensate contribution to the atom inverse propagator on the off-diagonal. This is the contribution familiar from Bogoliubov theory in the low density limit, and we see that our generalization to arbitrary density produces such a structure also at high density. Now we have to consider the effect of the new vertices. As argued above, we can restrict ourselves to diagrams carrying a single one of them. We focus on diagrams which renormalize the inverse atom propagator. These are tadpole diagrams. The diagrams renormalizing the diagonal entries must be $\mathcal O (\delta \theta^2)$ in order to ensure particle number conservation. The diagrams renormalizing the off-diagonal entries must involve one of the new vertices $\mathcal O (\delta \theta)$, and the trace over the inner line scales with a function $f(\delta \theta)$  with $f(0)=0$. Consequently, the fluctuation contributions are higher than linear order for both diagonal and off-diagonal entries and can be discarded. Thus, the full atomic mass matrix at $\mathcal O (\delta \theta)$ reads (we separate true potential (binding) energy from kinetic energy, $\mu (U) = E_b(U)/2 -Jz$ at $n\approx 0$,  $\mu (U) = - E_b(U)/2 + U + 2 Jz$ at $n\approx 2$ \cite{Diehl09I}, $z=2d$ the lattice coordination number)
\begin{widetext}
 \begin{eqnarray}\label{AtomGFull}
G^{-1}_{1,\theta_0=0} (\omega = 0;\mu , \textbf{q} = 0) &=&  \left(
\begin{array}{cc}
  {- E_b (U)/2 } & {2\sqrt{2} Jz \delta \theta/ 2} \\
  {2\sqrt{2} Jz \delta \theta/ 2} & {- E_b (U)/2}
\end{array}
\right), \\\nonumber
G^{-1}_{1,\theta_0=\pi} (\omega = 0; \mu , \textbf{q} = 0) &=&  \left(
\begin{array}{cc}
  { -E_b (U)/2 } & {-2\sqrt{2} Jz \delta \theta/ 2} \\
  {-2\sqrt{2} Jz \delta \theta/ 2} & { -E_b (U)/2 }
\end{array}
\right).
\end{eqnarray}
\end{widetext}
Hence we conclude that the dominant effect beyond mean field theory, which implies the simple linear relation $\mu(U) = U/2$ for the binding energy (see \cite{Daley09}, but the argument is repeated below Eq. \eqref{Crit2} for convenience), comes from vacuum fluctuations, which determine the value of $\mu (U)$ (or, equivalently, $E_b(U)$) as a function of the interaction strength $U$. These high energy fluctuations are responsible for shifts of the critical point. Note that the sign on the off-diagonal of the second equation is unphysical as it can be absorbed by a phase rotation of the order parameter. Thus, the equations have the same form.

The critical interaction strength can now be extracted from the characteristic equation  \eqref{characteristic} which reads explicitly, in both limits,
\begin{eqnarray}\label{lowcrit}
 (-E_b(U_c) /2)^2 - 2 (Jz \delta \theta)^2 = 0.
\end{eqnarray}

\label{sec:ASFDSFex}
\begin{figure}[t]
\begin{center}
\includegraphics[width=0.9\columnwidth]{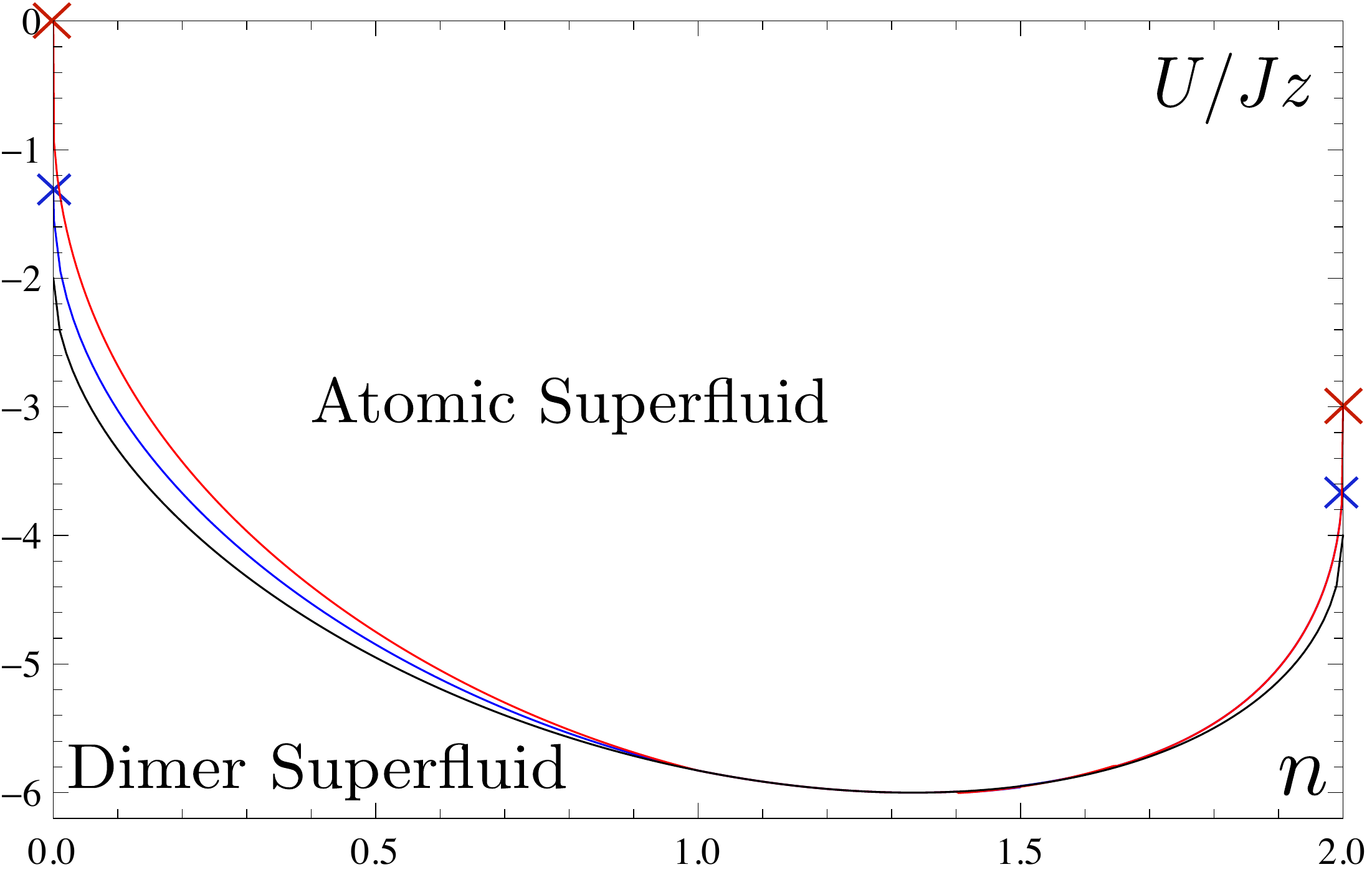}
\end{center}
\caption{\label{PhaseBorder} ASF-DSF phase boundary for the attractive 3-body hardcore constrained Bose-Hubbard model.  The black lower curve is the mean field result, while red upper and blue middle curve correspond to $d=2,3$. The crosses denote the endpoints of the critical lines. For high densities, the $d=2,3$ results are very close to each other except for a tiny region close to maximal filling. 
}
\end{figure}

The physical solution is given by $E_b(U_c) = - 2 \sqrt{2} Jz \delta \theta$, i.e. in the vacuum limit $\delta \theta\to 0$ also the binding energy vanishes. Thus, the critical interaction strength for the formation of the dimer or di-hole bound state coincides with the energy scale of the single particle excitations (atoms or holes) becoming critical. In these limits, we may quantitatively estimate the dependence of the interaction strength on e.g. the condensate fraction in two and three dimensions. In $d=3$, the binding energy starts quadratically due to the non-analyticity in the fluctuation integral. In contrast, in $d=2$, the fluctuation integral features the well known logarithmic behavior. This yields, in dimensionless units, the physical solutions
\begin{eqnarray}
n\to 0: \,\,d = 3 &:& \tilde U_c =  \tilde U_0 \big(1+ 2^{3/4} \sigma \sqrt{\,\,\delta\theta}\big), \\\nonumber
d=2   &:& \tilde U_c = 2\pi \big(\log \tfrac{4\sqrt{2} \delta \theta}{\Lambda^2} \big)^{-1} \\\nonumber
n\to 2: \,\,d = 3 &:& \tilde U_c =  \tilde U_2 \big(1+ 2^{-3/4} \sigma \tfrac{|\tilde U_2| -3}{|\tilde U_2|}\sqrt{\,\,\delta\theta}\big), \\\nonumber
d=2   &:& \tilde U_c = - 3 +\pi \big(\log \tfrac{2\sqrt{2} \delta \theta}{\Lambda^2} \big)^{-1}
\end{eqnarray}
with numbers $\tilde U_0 \approx -4/3, \sigma \approx 0.42, \Lambda \approx 5.50$, and $\tilde U_2 \approx -11/3$ \cite{Diehl09I}. We note that due to the formation of the di-hole bound state at a finite interaction strength despite the logarithmic dependence of the fluctuation integral, also the critical interaction strength for $n\to 2$ remains finite.

Importantly, we find a \emph{non-analytic} dependence of the critical interaction strength on the condensate density $\delta \theta$ which is due to the strong fluctuation effects in the vicinity of the bound state formation. This is in contrast to the mean field result, which shows a linear dependence on the condensate angle $\delta\theta = \sqrt{2n}$ (see \cite{Daley09} and Eq. (\ref{UIsing}) below).


\subsection{Phase Diagram}

The analysis of the low density limits in the last paragraph reveals that the strongest beyond mean field effect comes from the high energy vacuum fluctuations renormalizing the relation $\mu(U)$ away from its mean field value $\mu =  U/2$. In our practical implementation of the calculation of the phase boundary, we will rely on this separation of vacuum and many-body effects. For the calculation of the phase diagram at a fixed total density $n$, we now also take modifications of the mean field equation of state $n=2|s|^2$ into account. We discuss the equation of state \eqref{FullDens} in an approximation where we omit the three-point correlations from the outset. Furthermore, in our concrete computations, we restrict ourselves to the calculation of the atomic depletion $\langle b_1^\dag b_1\rangle$. We will find that this contribution is small compared to the condensate part and thus has a small influence on the phase border only -- the dominant effect shifting the border stems from the vacuum fluctuations leading to a strong modification of $\mu(U)$. Based on this observation, we do not expect that the dimer depletion strongly modifies the phase border. 

In order to calculate the atomic depletion, we first consider the quadratic spin wave theory, 
\begin{widetext}
\begin{eqnarray}
S_1[ b^\dag_1, b_1]  &=& - J \sum_{\langle i,j\rangle}\Big[ (c^2 + 2|s|^2) b_{1,i}^\dag b_{1,j} + \sqrt 2 c \,\, \big(s^* b_{1,i}  b_{1,j}  + s b_{1,i}^\dag   b_{1,j}^\dag \big)    \Big]  + [- \mu -(-2\mu +U)|s|^2)]  \sum_i b_{1,i}^\dag  b_{1,i} \\\nonumber
&=&  \frac{1}{2}\int _q  ( b_1^\dag (q) , b_1(-q)) \left(
\begin{array}{cc}
\mathrm i \omega + p_\bq   &\Delta_\bq \\
 \Delta_\bq & - \mathrm i \omega + p_\bq
\end{array} \right) 
\left(\begin{array}{c}
 b_1 (q)\\ 
 b_1^\dag (-q)
\end{array} \right) ,\\\nonumber
p_\bq &=& p_0 + \delta p_\bq, \quad p_0 = -\mu (1-2s^2) - U s^2 - Jz (1+s^2),\quad  \delta p_\bq = 2(1+s^2) \delta \epsilon_\bq ,\\\nonumber
\Delta_\bq &=& \Delta_0 + \delta \Delta_\bq, \quad \Delta_0 =  2\sqrt{2} Jz cs , \quad \delta \Delta_\bq = 4 \sqrt{2} cs \delta \epsilon_\bq ,\\\nonumber
\delta \epsilon_\bq &=& J \sum_\lambda (1 - \cos q_\lambda) .
\end{eqnarray}
\end{widetext}
The matrix in the second line is the inverse atomic Green function in frequency and momentum space $G_1^{-1}(\omega ;\mu, \bq )$. With these preparations, the approximate equation of state and the atomic depletion is found to be 
\begin{eqnarray}\label{Crit1}
n &=& 2 |s|^2  + (c - |s|^2)\langle b_1^\dag b_1\rangle ,\\\nonumber
 \langle b_1^\dag b_1\rangle &=& \Tr \,G_1(\omega;\mu, \bq )  = \frac{1}{2} \int\frac{d^dq}{(2\pi)^d}\Big(\frac{p_\bq}{\sqrt{p_\bq^2 - \Delta_\bq } } - 1\Big),
\end{eqnarray}
where we have performed the frequency integral by closing the contour in the upper half plane.

As stated above, we consider the renormalization effects on the inverse atom propagator which are present already in the vacuum problem. These are encoded in the value of the chemical potential $\mu (U)$ and depend on dimension. The condition determining $\mu$ reads 
\begin{eqnarray}\label{Crit2}
G^{-1}_{d/h}(\omega =0;\mu, \bq =0) =0,
\end{eqnarray}
where in the vicinity of $n=0$ we use the full dimer Green's function $G_d$, and close to $n=2$ the di-hole expression $G_h$ as given in Eq. \eqref{FullGreen}. In contrast, in the mean field approximation which uses the ``bare'' inverse dimer propagator $\mathrm i\omega -2\mu + U$, the above condition evaluates to $\mu (U)= -|U|/2$ independent of dimension and of whether we are close to zero or maximum filling. The critical point is determined by the atoms becoming unstable towards condensation. This is indicated by the condition 
\begin{eqnarray}\label{Crit3}
\det G_1^{-1}(\omega =0;\mu, \bq =0) = p_0^2  - \Delta_0^2 = 0.
\end{eqnarray}
We solve the system of equations (\ref{Crit1},\ref{Crit2},\ref{Crit3}) in two and three dimensions numerically. In particular, Eq. \eqref{Crit2} decouples from the other two equations within our approximation scheme yielding the renormalized relation $\mu(U)$, such that we merely need to solve (\ref{Crit1},\ref{Crit3}) with $\mu(U)$ as an input. The result for the phase diagram is plotted in Fig. \ref{PhaseDiagram}. We compare these results to the mean field approximation, which uses $\mu =U/2$ and $n=2|s|^2$ for the equation of state, resulting in the critical interaction
\begin{eqnarray}\label{UIsing}
\frac{U_c}{Jz}  
= -2 (\sqrt{1-n/2} + \sqrt{n})^2.
\end{eqnarray}
As anticipated above, the beyond mean field effects are mainly due to the fluctuations accompanying the bound state formation, which strongly modify the relation $\mu(U)$ as compared to the mean field value, while we find a subdominant role of the many-body depletion effects.The shape of the phase boundary directly reflects the non-analytic, dimension dependent behavior associated to the bound state formation. The quantitative effect is more pronounced below half filling than above. This may be traced back to the fact that the domain where nonperturbative fluctuation effects play a role is smaller in the high density regime than at small densities, cf. \cite{Diehl09I}. A simple picture can be given as follows: In the limits $n\to0,2$, the criteria for the atom criticality (zero eigenvalue of $G_1^{-1}$) and the microscopic bound state formation (zero eigenvalue of $G_{d/h}^{-1}$) coincide and fluctuation effects on the phase boundary are substantial. Moving away from these limits, the absolute value of the critical interaction increases, and the microscopic bound state is already tightly bound at the point where atom criticality is reached. The critical line then approaches the mean field phase boundary up to small perturbative corrections 
Thus, though our approximation is lacking a strict ordering principle when moving away from the limits $n\to0,2$, we therefore expect the mean field result to be rather accurate.
\begin{figure}[t]
\begin{center}
\includegraphics[width=0.9\columnwidth]{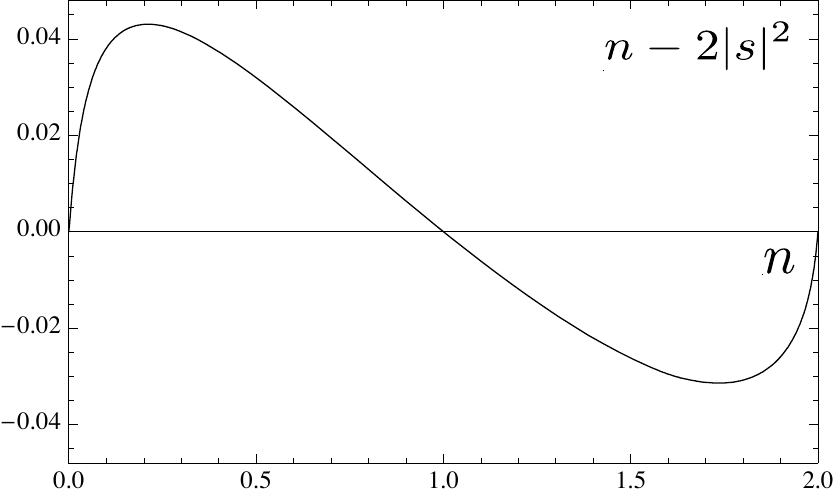}
\end{center}
\caption{\label{Depletion} Condensate depletion due to the atomic two-point function in two dimensions. The overall shape of the curve with zero crossing and sign change is determined by the function $c^2-|s|^2$ premultiplying the two-point function, cf. Eq. \eqref{FullDens}. The small overall size shows that the depletion effects produce only very tiny corrections to the phase border.}
\end{figure}

\section{Many-Body Physics in the Strong Coupling Regime and a Continuous Supersolid}
\label{sec:DSFCDW}

In this section we investigate the system in the strongly correlated limit $J/|U|\to 0$. In particular, we identify a bicritical point at half filling of atoms ($n=1$), at which homogeneous superfluid order (spontaneous phase symmetry breaking) and charge density wave order (spontaneous translation symmetry breaking) are degenerate. The bicritical point is due to a symmetry enhancement from the conventional $U(1) \simeq SO(2)$ to $SO(3)$, which is seen to be intimately connected to the 3-body constraint. We term the system in this regime a ``continuous supersolid'' due to the degeneracy of phase and translation symmetry breaking orders, where the order parameter may be rotated continuously from one to the other without energetic cost. This behavior is in contrast to other occurrences of supersolidity in bosonic systems \cite{SuSo}. Though this state is only reached asymptotically, it governs the physics in strong coupling and close to half filling, and we work out the observable consequences of this situation. We also propose a simple experiment to verify this scenario. 

\subsection{Analytical Approach}

Before embarking the calculation, let us stress that our beyond mean field approach is indispensable to settle these issues. Indeed, a straightforward comparison of the simple Gutzwiller mean field energies of the dimer superfluid and a charge density wave state (CDW) yields degenerate energies for these two states for all fillings: The Gutzwiller mean field CDW state is given by $|\text{CDW} \rangle =  \prod_{i\,\mathrm{even}} |2\rangle_{i}   |0\rangle_{i+1}$, and for a fixed average density $n$ has energy density $E_{\text{CDW}}/M^d = (U/2) n$, which precisely equals the mean field energy density $E_{\text{DSF}}/M^d$ of the dimer superfluid for all particle densities. In consequence, the question of the correct ground state cannot be decided within the simple Gutzwiller mean field theory (though a superfluid is clearly more plausible for incommensurate fillings). It is necessary to first integrate out the high energy single particle degrees of freedom, making the dimers true propagating and interacting physical excitations. Moreover, even the second order perturbation theory is not fully conclusive as we will see. The deviation from the second order result, calculated in \cite{Diehl09I}, plays a key role in the following discussion.

In \cite{Diehl09I} we have calculated the effective theory in the strong coupling limit: Perturbatively integrating out the single particle excitations up to fourth order, and taking the constraint principle for the effective action into account, we found the low energy effective Hamiltonian 
\begin{eqnarray}\label{Heffi}
H_{\mathrm{eff}} =  - t \sum_{\langle i,j\rangle}\big( t_{2,i}^\dag  X_iX_j t_{2,j} -  \lambda Ê\hat n_{2,i}\hat n_{2,j}\big)  -\mu_d  \sum_{i} \hat n_{2,i}  \nonumber\\
\end{eqnarray}
with $t = 2J^2/|U|$, $\mu_d$ the effective dimer chemical potential and the dimensionless ratio of nearest-neighbour interaction to hopping $\lambda =v/(2t)$ discussed below. Since in the perturbative limit there are only virtual single particle excitations, we may replace the constraint operator $X_i = 1-\hat n_{1,i}-\hat n_{2,i} \to 1-\hat n_{2,i}$. In this case we have the following mapping to effective spin $1/2$ degrees of freedom, which will more clearly reveal the physics of the model,  
\begin{eqnarray}\label{eq:spinmapping}
s_{j}^{+} &=&(s^x_{j}+is^y_{j})\equiv  (-)^j t_{2,j}^{\dag }X_{j}, \\\nonumber
s_{j}^{-} &=&(s^x_{j}-is^y_{j})\equiv (-)^j  X_{j}t_{2,j}, \\\nonumber
s^z_{j} &=&\hat n_{2,j}-1/2,
\end{eqnarray}%
where on the bipartite lattice with sublattices $A$ and $B$ we use $(-)^j = +$ for $j\in A$ and $(-)^j = -$ for $j\in B$. Up to a constant the Hamiltonian then takes the form 
\begin{eqnarray}\label{Heffective3}
H_{\mathrm{eff}}=2t\sum\limits_{\left\langle i,j\right\rangle }\left(s^x_{i}s^x_{j}+s^y_{i}s^y_{j}+\lambda s^z_{i}s^z_{j}\right)  - \mu_d \sum\limits_{i}s^z_{i}.
\end{eqnarray}
The anisotropy parameter $\lambda=v/(2t)$ evaluates to $\lambda =1$ in the second order perturbation theory, corresponding to an isotropic antiferromagnetic Heisenberg model -- note the sign change in $t$ due to the sublattice dependent sign in \eqref{eq:spinmapping}. The fourth order calculation yields  \footnote{One may wonder about implicit density effects for the perturbative calculation, that is, if the perturbative calculation at zero density is sufficient for the calculation of the effective theory for all densities. This may be discussed by studying the limits $n=0,2$. At second order, the emergent particle-hole symmetry ensures that the perturbative results coincide for both cases, while at fourth order, differences occur, pointing at the above mentioned implicit density effects. However, due to the identical diagrammatic structure one still finds $\lambda<1$ for $n=2$, which is the crucial ingredient for the argument presented here.}
\begin{eqnarray}
\lambda = 1 - 8 (z-1) \big(\frac{J}{U}\big)^2 <1 .
\end{eqnarray}
This result has been derived in \cite{Diehl09I}, cf. Sect. V.D.2, Eq. (60). It is obtained as the ratio of dimer-dimer interaction and dimer hopping coefficient calculated at fourth order. Next-to-nearest neighbour terms also appear at fourth order, but their numerical coefficient is much smaller than for the nearest neighbours due to the restricted pathways contributing to these terms, and are thus neglected.

For $\lambda =1$ and half filling of atoms $n=1$, where the term involving the chemical potential vanishes, the system exhibits a symmetry enhancement from $SO(2)\simeq U(1)$ (corresponding to rotations in the $x-y-$plane, or phase rotations,  generated by $\exp \mathrm i \theta_z S^z \propto \exp \mathrm i \theta_z \hat N$, with global operators $S^\alpha = \sum_i s_i^\alpha, \hat N = \sum_i \hat n_{2,i}$) to $SO(3)$ (corresponding to arbitrary rotations on the Bloch sphere $\exp \mathrm i \sum_\alpha \theta_\alpha S^\alpha$). 

The $SO(3)$-invariance of the quadratic part in \eqref{Heffective3} also implies a simple transformation behavior of the total Hamiltonian under a discrete particle-hole or charge conjugation transformation: The special choice $U=\exp (\mathrm i \pi S^{x})$ implements the mapping
\begin{eqnarray}
s_{j}^{\pm } \rightarrow U^{-1}s_{j}^{\pm }U=s_{j}^{\mp }, \quad s^z_{j} \rightarrow U^{-1}s_{j}^{z }U=-s^z_{j}.
\end{eqnarray}%
Under such a transformation $N\rightarrow M^d - N$ and, hence, $S^{z}\rightarrow M^d/2-S_{z}$. The particle-hole symmetry makes the phase diagram of deeply bound dimers symmetric under the replacement $\hat n_{2}\rightarrow 1-\hat n_{2}$. In general, such a symmetry is absent. Moreover, it is also not present in the opposite limit of strong repulsive interactions, which is asymmetric when $\hat n_{1}$ is replaced with $2-\hat n_{1}$.

The $SO(3)$-invariance is a peculiar feature of the leading order perturbation theory. Its physical origin is well understood in terms of the geometric argument which relates hopping and interaction paths, cf. Sec. IV D in the companion paper \cite{Diehl09I}. At second order, no other interaction processes can occur, and thus the hopping and interaction constants must be equal. However, as seen in \cite{Diehl09I}, at fourth order perturbation theory additional interaction processes yield $\lambda <1$, thus reducing the symmetry to $SO(2)$, and also spoiling the particle-hole symmetry. In addition, several other terms are generated, which describe next-to-nearest neighbour hopping and interaction, or three- and four-spin interactions. Nevertheless, the proximity to the Heisenberg point $\lambda=1$ has an impact on the phase diagram, and we will use its well know properties to understand the phase diagram and the nature of the low lying excitations in the perturbative regime. 

The order parameter for this model is given by the expectation value of the N\'eel vector $\hat{\mathcal{N}}^\alpha = \sum_j (-)^j s^\alpha_i$. Its vector character is under $SO(3)$,   $[S^\alpha , \hat{\mathcal{N}}^\beta] = \mathrm i \epsilon_{\alpha\beta\gamma} \hat{\mathcal{N}}^\gamma$, i.e. global spin rotations transform the N\'eel vector components into each other. Translating back to the original boson language, a finite $\langle\hat{\mathcal{N}}^z\rangle$ corresponds to charge density wave order, while finite values of $\langle\hat{\mathcal{N}}^x\rangle, \langle\hat{\mathcal{N}}^y\rangle$ indicate dimer superfluid (DSF) order. For the isotropic Heisenberg model without magnetic field (or at half filling), $[H,S^\alpha]=0$ for all $\alpha$, and thus CDW and DSF order are degenerate. The perturbative limit of our model thus realizes a bicritical point \cite{FisherNelson74} with two competing orders. Such an enhancement of internal symmetries is well known in magnetic systems \cite{Kohno97,Batrouni99}, but less common and intuitive in physically realizable bosonic models, which usually only exhibit phase symmetry. Due to the degeneracy of phase and translation symmetry broken states, both order parameters are generically nonzero, and we may term the state a continuous supersolid, whose experimental implications are studied below.

We can make this discussion even more explicit when changing from the spin to a hardcore boson language
\begin{eqnarray}
s_{j}^-   =(-1)^{j}h_{j},\,\,
s_{j}^+   =(-1)^{j}h_{j}^{\dag},\,\,
s_{j}^z   =h_{j}^{\dag}h_{j}-\tfrac{1}{2},
\end{eqnarray}
where the hardcore bosons obey $h^{\dag\,2}_{j}\equiv 0$. We consider infinitesimal $SO(3)$ transformation with the parameters $\varepsilon_{\alpha}\ll 1 (\alpha =(x,y,z))$, $\delta s_{j}^\alpha=i\left[  S,s_{j}^\alpha\right]$, where $S=\sum_{i\beta}\varepsilon_{\beta}s_{i}^\beta$. The explicit form of the
above transformation reads ($s_{j}^\pm=s_{j}^x\pm is_{j}^y$)  
\begin{eqnarray}
\delta s_{j}^+  & =&-i\varepsilon s_{j}^z -i\varepsilon_{z}s_{j}^+,\\\nonumber
\delta s_{j}^-  & =&i\varepsilon^{\ast}s_{j}^z +i\varepsilon_{z}s_{j}^-,\\\nonumber
\delta s_{j}^z  & =&-\tfrac{i}{2}\varepsilon^{\ast}s_{j}^+ +\tfrac{i}{2}\varepsilon s_{j}^-,%
\end{eqnarray}
with $\varepsilon=\varepsilon_{x}+i\varepsilon_{y}$. In terms of bosonic
hardcore operators one thus obtains
\begin{eqnarray}
\delta h_{j}^{\dag}  & =&(-1)^{j}\left[  -i\varepsilon(h_{j}^{\dag}%
h_{j}-\frac{1}{2})\right]  -i\varepsilon_{z}h_{j}^{\dag},\nonumber\\
\delta h_{j}  & =&(-1)^{j}\left[  i\varepsilon^{\ast}(h_{j}^{\dag}%
h_{j}-\frac{1}{2})\right]  +i\varepsilon_{z}h_{j},\nonumber\\
\delta(h_{j}^{\dag}h_{j}-\frac{1}{2})  & =&(-1)^{j}\left[  -\tfrac{i}%
{2}\varepsilon^{\ast}h_{j}^{\dag}+\tfrac{i}{2}\varepsilon h_{j}\right]  .
\end{eqnarray}
Note that the last terms in the first and second lines are just usual gauge transformations.

The change in the operators results in the change of their mean-field values.
If one introduces the usual superfluid order parameter $\psi=\sum_j\left\langle
h_{j}\right\rangle $ and the CDW order parameter $\mathcal{N}\equiv \langle \hat{\mathcal{N}}^z\rangle=\sum_j(-1)^{j}\langle (h_{j}^{\dag}h_{j}-\frac{1}{2})\rangle $, then the
corresponding change in the order parameters is%
\begin{eqnarray}
\delta\psi & =&i\varepsilon_{z}\psi+i\varepsilon^{\ast}\mathcal{N},\\\nonumber
\delta \mathcal{N}  & =&-\tfrac{i}{2}\varepsilon^{\ast}\psi^{\ast}+\tfrac{i}{2}%
\varepsilon\psi.
\end{eqnarray}
It is easy to check that the above transformation leaves the combination
$\mathcal{N}^{2}+\left\vert \psi\right\vert ^{2}$ invariant. We therefore can conclude
that the $SO(3)$ symmetry corresponds to canonical transformations of the dimer
operators, which change both superfluid and CDW order parameters, but leave
the combination $\mathcal{N}^{2}+\left\vert \psi\right\vert ^{2}$ invariant.

Another important point about the $SO(3)$ symmetry is that it is broken for
$n\neq1$ not on the Hamiltonian level (the Hamiltonian with $\lambda=1$ is
always $SO(3)$ symmetric) but on the level of the subset of states (with a
fixed $S^{z}$), on which it has to be minimized. In a generic case $n\neq1$
(and, hence, $S^{z}\neq0$) the subspace reduces the symmetry down to
$SO(2)\simeq U(1)$ gauge group. In the case $n=1$ with $S^{z}=0$, however, the
subspace contains the manifold of spin-singlet (and, therefore, $SO(3)$
symmetric) states, which have the lowest energy. The symmetry transformation
corresponds simply to the motion on this manifold.

A similar scenario (an enhancement to a pseudo $SU(2)$ symmetry) is actually observed in attractive lattice fermion systems \cite{Zhang90}. Similar to the fermion system, the symmetry enhancement is thus a unique consequence of the 3-body hardcore constraint. Indeed, attractive lattice bosons without such constraint, analyzed in detail by Petrosyan, Schmidt \emph{et al.} \cite{Fleischhauer06}, show a different behavior: Due to the possibility of virtually occupying a lattice site with three atoms, it is found $\lambda =4$. This places the unconstrained attractive bosons in the ``Ising limit'', which was analyzed further in the latter reference. 

As we find $\lambda <1$ in fourth order perturbation theory, the bicritical point is approached from the homogeneous superfluid, which is energetically favoured over the charge density wave. Nevertheless, we may expect important observable consequences. Indeed, the symmetry enhancement $SO(2)\to SO(3)$ implies the emergence of a second gapless, and therefore collective, Goldstone mode. For a weakly explicitly broken $SO(3)$ symmetry, one still has a near gapless collective mode with experimentally observable consequences. We propose an experiment, which is based on the idea of explicitly rotating the macroscopic N\'eel vector from the $x-y-$plane representing superfluidity to the $z-$axis, realizing a CDW ordered state. We will also show that this experiment allows to quantitatively characterize the pseudo Goldstone mode. 

To favor CDW ordering, we explicitly break the lattice translation symmetry via introduction of a superlattice shifting the single particle energies on adjacent sites:
\begin{eqnarray}
-\mu_d  \sum_i \hat n_{2,i} &\to& -\mu_A\sum_{i\in A} \hat n_{2,i} - \mu_B \sum_{j\in B} \hat n_{2,j}  \\\nonumber
&=& -\mu \sum_i \hat n_{2,i}  + \bar \nu\Big( \sum_{i\in A} \hat n_{2,i} - \sum_{j\in B} \hat n_{2,j}\Big) ,\\\nonumber
\mu &=&\frac{\mu_A + \mu_B }{2},\quad \bar\nu = \frac{\mu_A - \mu_B}{2}.
\end{eqnarray}
Here, $\mu$ is and average chemical potenial and $\bar\nu$ an imbalance parameter. Now we calculate the mean field ground state as well as the spectrum of excitations of the effective low energy Hamiltonian \eqref{Heffi}, using the rotation formalism (cf. Sec. \ref{sec:Review}). The approach is fully equivalent to the leading order $1/S$ expansion, which is not a well controlled expansion for $S=1/2$, but is known to yield the main features of the Heisenberg model in external fields. 

At the Heisenberg point, the $SO(3)$ spin symmetry requires an enlarged parameter space for the order parameter describing the ground state of the system. We consider an ansatz parameterized by two angles for the rotation of the $t_2$ degree of freedom (in contrast to the $U(1) \simeq SO(2)$ case with a single rotation angle for $t_2$), 
\begin{eqnarray}
R (\theta, \varphi_l) &=&  \left(
\begin{array}{cc}
  {\cos (\theta+\varphi_l)/2} &  {\sin (\theta+\varphi_l)/2\mathrm e^{2\mathrm i \phi}}\\
  {-\sin (\theta+\varphi_l)/2 \mathrm e^{-2\mathrm i \phi}} &  {\cos (\theta+\varphi_l)/2}
\end{array}
\right) .\nonumber\\
\end{eqnarray}
Here, $l=A,B$ is an index which depends on the sublattice $A$ or $B$, thus enabeling the description of a spatially modulated phase. For example, the choice $\theta =0, \varphi_A = 0,\varphi_B=\pi$ describes a charge density wave. The homogeneous choice $\theta\neq 0 , \varphi_A= \varphi_B=0$ describes a superfluid ground state. The rotation matrix is only $2\times2$ since we have integrated out the atoms. 

Expanding the thus rotated Hamiltonian, and replacing $b_{0,i}\to X_i = 1 - \hat n_{2,i}$, to second order we obtain
\begin{widetext}
\begin{eqnarray}
E_{\mathrm{mf}}/M^d &=& -2tz (s_A c_A s_B c_B - \lambda s_A^2s_B^2 ) - \frac{\mu}{2}(s_A^2 + s_B^2) + \frac{\bar\nu }{2}(s_A^2 - s_B^2) ,\\\nonumber
H_{\mathrm{lin}} &=& \,\,\,\,\big[ tz\big(- c_Bs_B(c_A^2 - s_A^2) + 2\lambda s_B^2 s_A c_A\big) + (- \mu + \bar\nu ) c_As_A\big] \sum_{ i\in A}(b_{2,i}^\dag +b_{2,i})  \\\nonumber
&& +\big[tz \big( - c_As_A(c_B^2 - s_B^2)+  2\lambda s_A^2 s_B c_B\big)  + (- \mu  - \bar\nu ) c_Bs_B\big]\sum_{ j\in B}(b_{2,j}^\dag + b_{2,j}),\\\nonumber
H_{\mathrm{SW}} &=&\frac{ t}{2}  \sum_{\langle i,j\rangle}  \big[\big(- (c_A^2c_B^2 + s_A^2s_B^2)-2 \lambda  c_A s_A c_Bs_B \big) ( b_{2,j}^\dag b_{2,i} + b_{2,i}^\dag b_{2,j}) 
\\\nonumber
&& \qquad \,\,+ \big(c_A^2s_B^2 + c_B^2s_A^2 - 2 \lambda  c_A s_A c_Bs_B \big)(b_{2,j} b_{2,i} + b_{2,i}^\dag b_{2,j}^\dag )  \big]
\\\nonumber
&&+ \big(2tz(2c_A s_A c_Bs_B -\lambda s_B^2(c_A^2 - s_A^2)) + (-\mu+\bar\nu )(c_A^2 - s_A^2) \big)\sum_{i\in A}  b_{2,i}^\dag b_{2,i} \\\nonumber
&&+\big(2tz(2c_A s_A c_Bs_B -\lambda s_A^2(c_B^2 - s_B^2)) + (-\mu-\bar\nu ) (c_B^2 - s_B^2)\big)\sum_{j\in B}  b_{2,j}^\dag b_{2,j}  . 
\end{eqnarray}
\end{widetext}
Here $s_l = \sin (\theta+\varphi_l)$ etc., and we have set the spontaneously chosen phase $\phi=0$ without loss of generality. $M$ is the total number of sites in each lattice direction. Since we break the lattice translation symmetry via our choice of the ansatz for the ground state, it is important to be careful with the position indices -- $b_{2,i}$ are located on the sublattice $A$, $b_{2,j}$ on the sublattice $B$. Eventually we are interested in a situation at fixed density. The local density operator to quadratic order takes the form
\begin{eqnarray}
t_{2,i}^\dag t_{2,i} = s_l^2 + c_ls_l (b_{2,i}^\dag  + b_{2,i}) + (c_l^2 - s_l^2 ) \hat n_{2,i},
\end{eqnarray}
where $l=A(B)$ for $i \in A(B)$. Thus, in the mean field approximation the equation of state reads
\begin{eqnarray}
n = N/M^d = \tfrac 2 2 (s_A^2 + s_B^2).
\end{eqnarray}
At half filling $n=1$, this implies $s_A = c_B, s_B = c_A$. Together with the relations $s_l^2 + c_l^2  = 1$, within this approximation everything may be expressed in terms of e.g. $s_A$ alone. In particular, the mean field energy determining the ground state takes the form 
\begin{eqnarray}
E_{\mathrm{mf}}/M^d &=& -2tz s_A^2(1- s_A^2)(1 - \lambda ) +\bar\nu (2s_A^2 -1).
\end{eqnarray}
The ground state is found from identifying the stable minima with respect to variation in $\theta$, and thus we have dropped the contribution from the chemical potential, as it contributes a rotation angle independent constant only for effectively fixed density. For $\bar\nu=0$, the ground state for $\lambda <1$ is the homogeneous superfluid with $s_A^2=1/2$. For $\lambda>1$, the charge density wave with $s_A=0,s_B=1$ is favored. At the Heisenberg point $\lambda=1$, both states are degenerate in accord with the exact symmetry argument. Now we consider the relevant case $\lambda <1$. Tuning $\bar\nu$ away from zero by ramping the superlattice, the superfluid acquires a spatial modulation, $s_A^2 = (1 + \nu)/2, s_B^2 = (1-\nu)/2, \nu = \bar\nu/tz(\lambda-1)$, where at a critical value 
\begin{eqnarray}
|\nu_c| =1, \quad |\bar\nu_c| = tz(1-\lambda) \approx 8tz(z-1)(J/U)^2
\end{eqnarray}
the SF is destroyed in favor of the CDW. Thus, ramping the superlattice corresponds to rotating the N\'eel order parameter. As we will see below, the critical value corresponds precisely to the gap of the pseudo-Goldstone mode. Hence, via measurement of the SF correlations \cite{Altman04}, which cease to exist at $\bar\nu_c$, one can quantitatively determine the characteristic property of the additional collective mode.  

The chemical potential $\mu$ is determined from the equilibrium condition that the linear terms vanish, evaluating to $\mu = \lambda tz$ independent of $\bar\nu$. Inserting this and the above expression for $s_A$, and switching to the Lagrange formalism, we obtain the Gaussian action
\begin{widetext}
\begin{eqnarray}\label{GaussAct}
S &=& \frac 1 2 \int_{q,q'} \delta (q -q') (b_2(-q), b_2^\dag (q)) 
\left(\begin{array}{cc}
g_\bq & h_\bq (-\omega) \\
h_\bq (\omega) & g_\bq 
\end{array}\right)
\left(\begin{array}{c}
b_2(q') \\
b_2^\dag (-q')
\end{array}\right),\\\nonumber
g_\bq &=& tz  ( \tfrac{\lambda -1}{2}(1-\nu^2) +1) \tilde \epsilon_\bq , \quad h_\bq (\omega) = \mathrm i \omega  + tz (1  +   \tfrac{\lambda - 1}{2} (1-\nu^2)\tilde \epsilon_\bq ),
\quad \tilde \epsilon_\bq = \tfrac 1 d \sum_\lambda \cos q_\lambda.
\end{eqnarray}
\end{widetext}
The spectrum of excitations may be computed from the condition that the determinant of the fluctuation matrix vanish. We obtain  
\begin{eqnarray}
\omega (\bq) = \pm tz\big( ((1+ [\lambda -1][1-\nu^2] )\tilde \epsilon_\bq +1 ) (1-\tilde \epsilon_\bq ) \big)^{1/2} .
\end{eqnarray}
For $\bar\nu=0$ and $\lambda <1$, the dispersion simplifies to  $\omega (\bq) = \pm tz\big( (\lambda\tilde \epsilon_\bq +1 ) (1 - \tilde \epsilon_\bq ) \big)^{1/2}$, and there is a single Goldstone mode at $\bq =0$, corresponding to the spontaneously broken $U(1)\simeq SO(2)$ symmetry in the dimer superfluid. For $\lambda \lesssim 1$, there is a second near gapless mode with gap $tz (1-\lambda)$ located at the edges of the Brillouin zone $\bq = \bf{\pi}$. At the Heisenberg point $\lambda=1$, this gap closes, and the system features the two Goldstone modes corresponding to the spontaneously broken $SO(3)$ symmetry. The system is then at the bicritical point where the order parameter can be freely rotated on the Bloch sphere. In the general case, the gap of the second near gapless mode is given by 
\begin{eqnarray}
\Delta = tz ( 1- \lambda )(1-\nu^2) ,
\end{eqnarray}
and we observe that we reach a point where there are two exactly gapless modes by tuning $\bar\nu\to \bar\nu_c, \nu^2 \to 1$. In this case, the two gapless modes correspond to the characteristic excitations on an antiferromagnetic, or CDW, ground state, and there is no superfluid order as the Bloch vector is confined to the $z-$axis. In Tab. \ref{Dispersions}, we summarize the dispersions found in the different density regimes in the leading order perturbation theory limit $\lambda =1$, and $\bar\nu=0$.

In summary, we propose a conceptually simple experiment that allows to rotate the macroscopic N\'eel vector order parameter via ramping a superlattice. The measurement of superfluid and density-density correlations \cite{Altman04} allows to monitor this rotation, as well as to measure the gap of the collective pseudo-Goldstone mode, which is the hallmark of the proximity of the system to the bicritical point with enhanced symmetry. Alternative experiments for the investigation of this proximity include a direct measurement of the dispersion relation via Bragg spectroscopy on the lattice \cite{Sengstock09}, or analyzing the system subject to slow rotation, which also acts as a current defavoring SF against CDW order \cite{Burkov08}.

We further comment on the relation of our spatially modulated superfluid for nonvanishing $\bar\nu$ to a supersolid. The latter is defined as a state with simultaneously and spontaneously broken phase and translation symmetry. In our case, both symmetries are broken, but the translation symmetry breaking is explicit and not spontaneous. Though the correlations are those of a supersolid, we would not term the state as such. 

Finally, we note that the evolution of the system from repulsive to attractive coupling may be viewed as a transition from a spin 1 model (3 onsite states) to a spin 1/2 model. The x-y ordered phases of these two models are separated by the Ising transition discussed in more detail in the next section.

\begin{table}[h] \caption{\label{Dispersions} Low energy dispersions in the perturbative regime (second order). $z$ is the dynamic exponent, $\omega \sim |\textbf{q}|^z$. }
  \begin{ruledtabular}
    \begin{tabular}{|c|c|c|c|c|c|c}
    & $n=0$ & $0<n<1$ & $n=1$ & $1<n<2$ &  $n=2$ &\\\hline
     $z$ & $2$ & $1$  & $1$ & $1$ &  $2$ &  \\\hline
    $\#$ zero modes &  $1$ & $1$  & $2$ & $1$ & $1$ &    \\
    \end{tabular}
  \end{ruledtabular}
\end{table}

\subsection{Complementary exact numerical study in one dimension}

\begin{figure}[tb]
\begin{center}
\includegraphics[width=0.9\columnwidth]{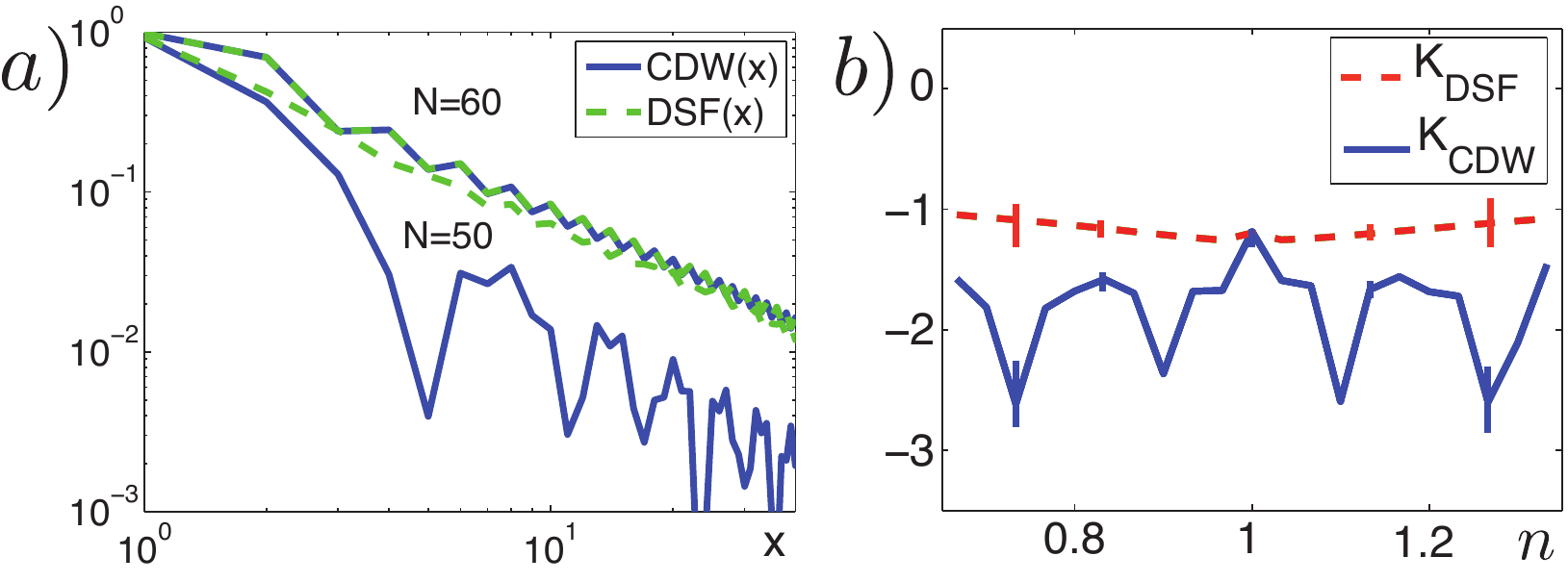}
\end{center}
\caption{Computations of the ground state on a 1D lattice with 60 sites for $U/J=-20$ using TEBD methods. (a) Correlation functions characterizing the CDW and DSF phases in a system with open boundary conditions ploted on a log-log scale as a function of distance $x$. Values are shown for $N=60$ particles, where the correlation functions are almost identical, and $N=50$ particles, where the decay of the CDW correlations are significantly more rapid than the DSF correlations (b) Algebraic decay exponents $K_{DSF}$ and $K_{CDW}$ that are fitted to the envelope of the correlation functions for varying mean density $n$. Error bars show typical errors in the fitted decay.}
\label{numfig1}
\end{figure}

We now investigate how the key features of these results manifest themselves in a 1D system. This can be done by computing the ground state of the constrained Bose-Hubbard model using the Time Evolving Block Decimation (TEBD) algorithm \cite{tebdvidal}
. Note that we optimise our algorithm for the conserved total number of particles \cite{daley05}, analogously to the optimisation for good quantum numbers in Density Matrix Renormalisation Group methods \cite{tdmrgdaley}
. In Ref. \cite{Daley09} we already observed quasi off-diagonal long range order in the Dimer Superfluid (DSF) correlation function $\langle b_i^\dag b_i^\dag b_{i+x} b_{i+x} \rangle$, together with exponential decay of off-diagonal elements in the single-particle density matrix $\langle b_i^\dag b_j \rangle$. This indicated the transition between the ASF and DSF phases in the 1D system.

\begin{figure}[tb]
\begin{center}
\includegraphics[width=0.9\columnwidth]{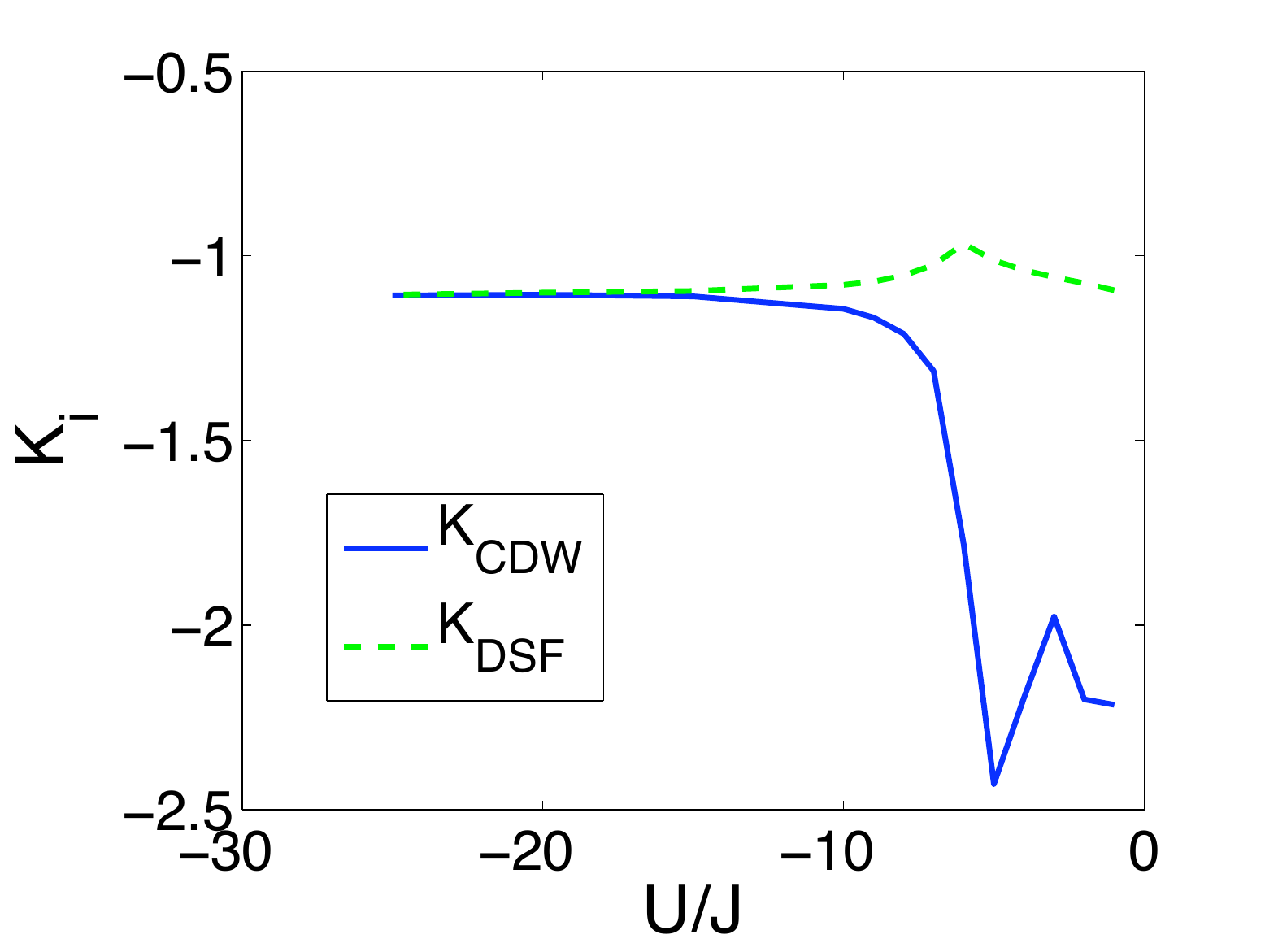}
\end{center}
\caption{Numerical validation of symmetry enhancement $SO(2)\to SO(3)$. The plot shows the exponents $K_{DSF}$ and $K_{CDW}$ describing the algebraic decay of DSF and CDW order in the ground state as a function of the ratio $U/J$ at unit filling $n=1$. For sufficiently large interactions, the coincidence of the decay exponents signals the degeneracy of the two kinds of order. These calculations were performed with TEBD methods for 60 particles on 60 sites. Open boundary conditions were used, but the decay exponents were fitted within the central 30 sites. The fitting errors are similar to those in Fig.~\ref{numfig1} (b). Number of Schmidt coefficients retained in TEBD calculations $\chi=200$.}
\label{numfig1B}
\end{figure}

Here we particularly investigate the situation near half-filling $n=1$, paying attention to the interplay between DSF order and CDW order, characterized by the density-density correlation function $CDW(x)=\langle n_i n_{i+x} \rangle -\langle n_i \rangle \langle n_{i+x} \rangle$. In Fig.~\ref{numfig1} we compare the DSF and CDW correlation functions for the ground state on a 60 site lattice with $U/J=-20$ and open boundary conditions. In Fig.~\ref{numfig1}a we plot the correlation functions both for $N=60$ (half filling of dimers) and $N=50$. At half filling the algebraic decay of these correlation functions is essentially the same, indeed the correlation functions are essentially equal, indicating coincidence of CDW and DSF orders in this state. Whilst reducing the total number of particles on the lattice to $N=50$ does not significantly change the DSF, the density-density correlation function decays much more rapidly in the ground state, in addition to large superimposed oscillations. This relative sensitivity of the correlation functions is characterised in Fig.~\ref{numfig1}b, where we show the result of fitting an algebraic decay $x^{K_i}$ to the envelope of each of the correlation functions. Again, we see that the decay of CDW and DSF correlations is identical within fitting errors at unit filling, but the CDW is very sensitive to deviations from unit filling, and it is dominated away from $n=1$ by the DSF. 

In Fig. \ref{numfig1B}, we study the approach of the bicritical point at fixed half filling $n=1$ as a function of the ratio of hopping and interaction $J/U$. The plot clearly shows the symmetry enhancement from the conventional $U(1)\simeq SO(2)$ to an $SO(3)$ symmetry: The two decay exponents describing DSF and CDW order approach each other for sufficiently strong attractive onsite interaction $U$, thus indicating the degeneracy of the two different kinds of order. 

\begin{figure}[tb]
\begin{center}
\includegraphics[width=0.9\columnwidth]{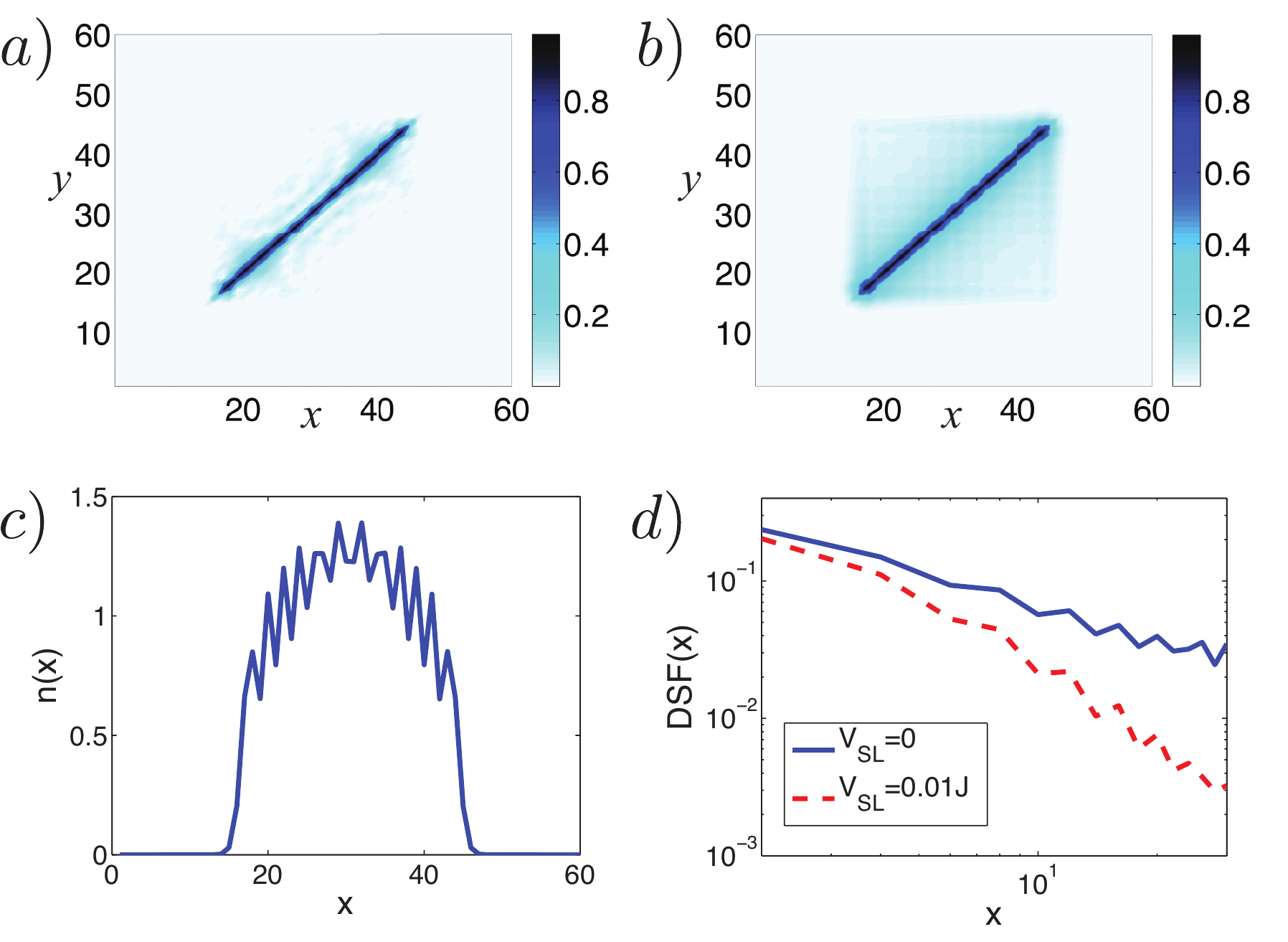}
\end{center}
\caption{Computations of the ground state with 30 particles on 60 lattice sites with $U/J=-20$, in the presence of a harmonic trap with on-site potential $V(x)$. (a) Shaded plot of the CDW correlation function $\langle n_x n_y \rangle -\langle n_x\rangle \langle n_y \rangle $ (with interpolated colours, and diagonal elements not shown), showing substantial order near $x=y=20$ and $x=y=40$, where the mean filling factor $n\sim 1$ regions. $V(x)= V_{\rm tr}(x-30.5)^2$, $V_{\rm tr}=1.33\times 10^{-3}$. (b) Shaded plot of the DSF correlation function $DSF(x-y)= \langle b_x^\dag b_x^\dag b_y b_y\rangle$ (with interpolated colours), showing substantial order across the occupied region of the lattice, for the same trap parameters as part a. (c) Density $n(x)$ for the same parameters as parts a,b. (d) Plots of the DSF correlation function $DSF(x)$ for a trap $V(x)=V_{\rm tr}(x-30)^2$, $V_{\rm tr}=0.7/30^2\approx 7.78 \times 10^{-4}$, with and without an additional superlattice potential $V_{SL}/2\sum_{i} (-)^i $.}
\label{numfig2}
\end{figure}

In an experiment it would be difficult to produce a setup with an exactly commensurate number of particles and lattice sites. One way to observe emergence of the CDW order, though would be to prepare the system in a harmonic trapping potential, where the density would vary across the trap. In Fig.~\ref{numfig2} we investigate the ground state for 30 particles on 60 lattice sites in the presence of such an external harmonic trap. In Figs.~\ref{numfig2}a,b we show the correlation functions for CDW and DSF order as they vary across the trap. We note that the DSF order is significant throughout the occupied region. We have checked in addition that across this region, the off-diagonal elements decay algebraically as a function of distance. On the other hand, the CDW correlations are most significant in regions near unit filling. For the trap parameters chosen here, this occurs near sites $20$ and $40$, as shown in Fig.~\ref{numfig2}c, where we also see significant oscillations in the density, which are also characteristic of the appearance of CDW order. In Fig.~\ref{numfig2}d we then investigate how the order can be manipulated by the addition of a weak superlattice. We see that the addition of an alternating potential on the order $0.01 J$ is sufficient to significantly increase the algebraic decay exponent for DSF order. Because the system size is small, it was difficult to obtain reliable results for the algebraic decay exponent of CDW order, but our calculations indicate that applying such a superlattice can indeed be used to select the dominant order for a system in the presence of a harmonic trap. 

Using t-DMRG methods we can also investigated possible time-dependent preparation of the continuous supersolid beginning from a Mott insulating state in the presence of a superlattice, analogously to the studies performed in Ref.~\cite{Daley09}. Beginning in an insulating state with two atoms in the lowest wells of a period two superlattice, it is possible to prepare a state with $n=1$ and $U/J=-20$ in a timescale of the order of $100J^{-1}$, with good fidelity of the DSF correlation functions, provided that a sufficiently strong constraint can be imposed so that no loss events occur on the timescale of the ramp.

\section{Long Wavelength Limit: Nature of the Phase Transition}
\label{sec:IR}

Even at low energies, non-linearities in the effective action may in principle have an impact on the physical observables, such as the nature of the phase transition. Such a scenario is known as Coleman-Weinberg phenomenon \cite{Coleman73}: Two near gapless degrees of freedom are coupled to each other, in a way that a phase transition which one of them undergoes is driven first order due to the long wavelength fluctuations of the other: the first order transition is radiatively induced. 

In our problem, indeed we face competing low energy degrees of freedom at the ASF-DSF transition: First, there is the gapless Goldstone mode present in the dimer condensate, which does not undergo qualitative changes at the ASF-DSF transition point. Second, at the Ising type transition one expects a $Z_2$ degree of freedom to emerge in the low energy sector for the atom degrees of freedom. A possible coupling between those low energy degrees of freedom may or may not give rise to a Coleman-Weinberg mechanism. 

Here we study this question by means of a systematic derivative expansion of the effective action. At ``low'' densities $n\approx 0,2$ we identify a first order transition in line with known results for continuum bosonic Feshbach models at low density \cite{Radzihovsky04,Sachdev04,Radzihovsky07}. Such a reproduction of the continuum results must be generally expected in low density lattice systems. This situation is seen to be rather generic in nonrelativistic systems \cite{Balents97,Lee04,Vojta00}. But, intriguingly, there is a lattice based decoupling mechanism which guarantees the existence of a second order transition, and thus a true quantum critical point, in the vicinity of $n=1$. Thus, we identify a true Ising quantum critical point in our system, connecting the two ordered ASF and DSF phases. 

Note that the scenario crucially hinges on the control over a coupling of the two near gapless modes close to the transition. It is evident that the discussion cannot be lead based on a simple quadratic spin wave theory.

\subsection{Low Energy Derivative Expansion}
\label{sec:DerivExp}

Our strategy is as follows: We will approach the phase transition from the DSF side, where there is not yet an atomic condensate, and tune the atomic mass parameters to criticality from there. For this purpose, we draw the low energy, continuum limit of the effective action corresponding to Eq. \eqref{HtNew}. We then identify the relevant low energy fluctuations and integrate out the massive degrees of freedom. We arrive at an action that describes the dimer Goldstone physics, the Ising degree of freedom as well as a cubic coupling of Goldstone mode to Ising density. The derivation is similar to the one presented in \cite{Radzihovsky07} in the continuum, differs however in the crucial aspect that the model discussed there features already microscopic propagating dimer degrees of freedom. Here we show how such terms are generated via successive integration of the massive degrees of freedom. 

At low energies, the action corresponding to the Hamiltonian \eqref{HtNew} encounters two immediate simplifications: First, we consider the constraint $X_i = 1- \hat n_{1,i}-  \hat n_{2,i}$: The density operators are less relevant than the number 1 at low energy. Consequently we replace $X_i \to 1$. By this replacement, we effectively drop the local constraint for the atoms and dimers. Physically, this is justified from the fact that infrared fluctuations with wavelengths much larger than the lattice spacing do not resolve single sites -- as stated above, while symmetries provide scale independent restrictions on the form of the effective action, the relevance of the constraint principle depends on scale. Second, we draw the continuum limit. Our original Hamiltonian \eqref{HtNew} often contains bilocal terms. In the quadratic sector, the resulting spatial derivative terms are kept: they describe spatial propagation and may be leading in the infrared for zero mass terms encountered close to the phase transition. However, in the interaction terms we drop the gradient couplings if they appear  in combination with a local one, which in comparison is always more relevant in the sense of the renormalization group. 
Finally, we drop the local quartic terms coupling atoms with dimers, which are subleading in comparison with the local cubic ones. The corresponding action reads
\begin{widetext}
\begin{eqnarray}\label{LowEnAction}
S &=& S_1[ b^\dag_1,b_1] + S_2[b_2^\dag,b_2] + S_{\text{int}}[b^\dag_1, b_1,\sigma ,\pi] ,\\\nonumber
S_1[b_1] &=&
\int _x  b_{1}^\dag  [\partial_\tau  - \mu -(-2\mu +U)s^2)- J(1+|s|^2) (z + \triangle) ]   b_{1}- \sqrt{2} J cs\big(  b_{1}(z + \triangle) b_{1}  + c.c.  \big),\\\nonumber
S_2[b_2] &=& \int _x b_{2}^\dag [\partial_\tau + (-2\mu +U)(c^2-s^2)]   b_{2},\\\nonumber
S_{\text{int}}[b_1,b_2] &=& \int _x Jz \big[ 3 cs \,\, \big( b_{2}^\dag +  b_{2}\big)\hat n_{1}   - \sqrt{2}  (c^2 - s^2) \big( b_{2}^\dag b_{1}^2  +c.c.  \big) 
\big].
\end{eqnarray}
\end{widetext}
($\int _x= \int d \tau d^d x, x= (\tau, \vec x), \triangle$ the Laplace operator. We omit the $(\tau,\vec x)$ dependence of the field for brevity.) Here and in the following we have chosen $s$ real without loss of generality. 

In the next step we identify the relevant phase fluctuations. The terms in the action (\ref{LowEnAction}) are seen to be in two classes: The first one is made up of field combinations which transform according to $U(1)\times U(1)$ (the first phase rotation acts on $b_1$ and the second on $b_2$), i.e. they do not lock the phases. In contrast, the cubic interaction terms in the last line of Eq. \eqref{LowEnAction} lock the phases such that the residual symmetry is a single $U(1)$. (Such a mechanism breaking $U(1)\times U(1) \to U(1)$ is a consistency check for our theory, which emerges from a constrained version of the Bose-Hubbard model, in turn only possessing a single $U(1)$ symmetry.) The dominant temporal and spatial phase fluctuations thus originate from the vicinity of the phase constraint emerging from the phase locking of the atomic to the dimer phase, $\theta_{2}(x) = 2\theta_{1}(x)$. To bring out the physics of these fluctuations, it is convenient to perform a local gauge transformation on the $b_1$ field such as to absorb the $\theta_2$ fluctuations \cite{Radzihovsky07}. Here we work in cartesian coordinates for the fluctuating fields, and consequently the gauge transformation is realized linearly. The gapless phase fluctuations of the dimer field are represented by its imaginary part, $b_{2}(x)  = (\sigma (x) + \mathrm i \pi (x))/\sqrt 2$ (cf. Eq. \eqref{HermField}). To absorb the phase fluctuations into $b_1$, we introduce dressed fields according to 
\begin{eqnarray}\label{kappadef}
b_{1}(x) \to \tilde b_{1}(x) = b_{1}(x) (1  - \mathrm i \kappa\, \pi (x) ), \\\nonumber \kappa = (c^2 -s^2)/(2\sqrt{2}cs) . 
\end{eqnarray}
Now the gauge transformed action can be calculated. In this expression, we only keep leading terms which are affected at linear order in the infinitesimal rotation. The result is
\begin{widetext}
\begin{eqnarray}
S_1[\tilde b^\dag_1,\tilde b_1]  &=& \frac{1}{2}\int _x  (\tilde b_1^\dag ,\tilde b_1) \left(
\begin{array}{cc}
\partial_\tau   +m_1^2 - J(1+s^2) \triangle & -2\sqrt{2} J cs(z  + \triangle) \\
 -2\sqrt{2} J cs( z + \triangle) & - \partial_\tau   +m_1^2 - J(1+s^2) \triangle 
\end{array} \right) 
\left(\begin{array}{c}
\tilde b_1\\ 
\tilde b^\dag_1
\end{array} \right), \\\nonumber
S_{\text{int}}[\tilde b^\dag_1,\tilde b_1,\sigma ,\pi] &=& \int _x\sqrt{2} i \kappa  \partial_\tau \pi  
\tilde b_{1}^\dag \tilde b_{1} + Jz  \sigma \Big[ 3 \sqrt{2} c s\,\,\tilde b_{1}^\dag\tilde  b_{1}  -  (c^2 - s^2)  \big(\tilde b_{1} \tilde b_{1}  + \tilde b_{1}^\dag \tilde b_{1}^\dag \big) \Big]
\end{eqnarray}
\end{widetext}
with $m_1^2 \approx  |U|/2 -Jz (1+s^2)$, using $-\mu\approx |U|/2$ -- as we are only interested in the low energy limit, the precise value of the couplings is unimportant, and we will work with the mean field values (which are, however, expected to be rather accurate except for the small density regime $n\approx 0$, cf. Sec. \ref{sec:ASFDSF}).
As a preparation for the elimination of the massive modes, we further introduce hermitian fields for the single particle excitations $\tilde b_1 = \varphi + \mathrm i \psi$, such that the action reads%
\begin{eqnarray}
S_1[\varphi,\psi]  
= \frac{1}{2}\int _x  (\varphi ,\psi)\hspace{-0.1cm} \left(
\begin{array}{cc}
  m_+^2 - \xi_+^2\triangle & \mathrm i \partial_\tau \\
- \mathrm i \partial_\tau  &   m_-^2 - \xi_-^2\triangle
\end{array} \right) \hspace{-0.15cm}
\left(\begin{array}{c}
\varphi\\ 
\psi
\end{array} \right) ,\nonumber\\\nonumber
\hspace{-0.6cm}S_{\text{int}}[\varphi,\psi,\sigma,\pi]\hspace{-0.1cm} =\hspace{-0.2cm} \int_x \mathrm i \tfrac{\kappa}{\sqrt{2}}\partial_{\tau}\pi(\varphi^{2}+\psi^{2})+\sigma
(\lambda_{-}\varphi^{2}+\lambda_{+}\psi^{2}),\\\hspace{-0.4cm}
\end{eqnarray}
where $\xi_{\pm}^{2}=J(c\pm\sqrt{2}s)^{2}$, $m_{\pm}^{2}=-|U|/2-z\xi_{\pm}^{2}$,
$\lambda_{\pm}=Jz(3cs/\sqrt{2}\pm(c^{2}-s^{2}))$, and again we keep only leading
terms. As appropriate for the phase mode, the field $\pi$ interacts with the atomic
fields $\varphi$ and $\psi$ only through its time derivative, while the field
$\sigma$ interacts directly. Note that $m_{+}^{2}>m_{-}^{2}$, and upon
approaching the phase transition, $m_{+}^{2}$ hits zero prior to $m_{-}^{2}$
\cite{Radzihovsky04,Radzihovsky07}. Indeed the condition $m_+^2 =0$ coincides with Eq. \eqref{UIsing} if we also use the mean field equation of state $n=2s^2$. For vanishing $m_+$, we then find $m_-^2\approx  4\sqrt{2}cs Jz = 4\sqrt{n(1-n/2)} Jz$ within the mean field approximation for the high energy physics. Hence, the field $\psi$ (the imaginary part of the atomic field $\tilde{b}_{1}$) remains massive for any density $0<n<2$ at the transition, and we may safely integrate it out perturbatively at the one-loop level, while the remaining degree of freedom $\varphi$ becomes soft and plays the role of an Ising field. The resulting effective action for the fields $\varphi$, $\pi$, and $\sigma$ reads%
\begin{eqnarray}
S[\varphi,\pi,\sigma] \hspace{-0.1cm} & =& \hspace{-0.2cm}\int_x \Big\{\frac
{1}{2}\sigma(M^{2}-\xi_{\sigma}^{2}\Delta)\sigma 
+\mathrm i\sigma
\partial_{\tau}\pi +\mathrm i\frac{\kappa}{\sqrt{2}}\partial_{\tau}\pi\varphi^{2}\nonumber \\
&& \qquad +\zeta(\partial_{\tau}\pi)^{2}-\xi^{2}\pi\Delta\pi\nonumber\\ 
&&\qquad  +\frac{1}{2}\varphi(m_{+}^{2}-Z_{\varphi}\partial_{\tau}^{2}-\xi_{+}%
^{2}\Delta)\varphi \Big\},
\end{eqnarray}
where $M^{2}\sim\lambda_{-}^{2}/m_{-}^{2}$, $\zeta\sim\kappa^{2}/m_{-}^{2}$,
$\xi^{2}\sim\xi_{\sigma}^{2}\sim\lambda_{-}^{2}\xi_{-}^{2}/m_{-}^{4}$, and
$Z_{\varphi}\sim m_{-}^{-2}$. (Note that in the limit $\left\vert U\right\vert
\gg Jz$ both fields $\varphi$ and $\psi$ are massive and, after integrating
them out perturbatively, we get Eq. \eqref{Heffi} for the effective dimer Hamiltonian of the $b_{2}$ field.) The field $\sigma$ now becomes massive and can be integrated out as well. The final
effective action for the fields $\varphi$ and $\pi$ is%
\begin{eqnarray}\label{IRSeff}
S_{\mathrm{eff}}[\varphi,\pi]  & =\int_x\Big\{\frac{1}%
{2}\varphi(-Z_{\varphi}\partial_{\tau}^{2}-\xi_{+}^{2}\Delta+m_{+}^{2}%
)\varphi+\lambda\varphi^{4}\nonumber\\
& +\frac{1}{2}\pi(-Z\partial_{\tau}^{2}-\xi^{2}\Delta)\pi+\mathrm i\frac{\kappa}%
{\sqrt{2}}\partial_{\tau}\pi\varphi^{2}\Big\},
\end{eqnarray}
with $Z\sim M^{-2}$ and $\lambda\sim\lambda_{+}^{2}/M^{2}$. This action
describes a coupled theory for the Goldstone mode $\pi$ and the Ising mode
$\varphi$. Note that here we also keep a fourth order Ising coupling. Its presence being rooted in the tree-level $\sigma$ exchange, this coupling is positive. Thus, the low energy theory contains an Ising part,
i.e. a real field with quartic potential which exhibits $Z_{2}$ symmetry
breaking when $m_{+}^{2}$ turns negative. If this part of the action were
isolated, the transition would be in the Ising universality class, and
therefore of second order. In the presence of the Goldstone-Ising coupling,
more care needs to be taken: In general, a coupling of two (near) gapless
real bosonic degrees of freedom can lead to a fluctuation induced first order
phase transition, known as the Coleman-Weinberg phenomenon \cite{Coleman73}.
The Ising self-interaction $\lambda$ and the Ising-Goldstone coupling $\kappa$
can be compared via naive power counting \footnote{The power counting applied here is based on the effective relativistic Ising and Goldstone low energy actions with dynamical exponent $z=1$, and not the original nonrelativistic theory.}: the canonical dimension of $\lambda$
is $3-d$, and that of $\kappa$ is $(3-d)/2$. Thus, in any dimension the
corresponding terms have the same degree of relevance and therefore compete
with each other.

The form of the action \eqref{IRSeff} coincides with the one obtained in \cite{Radzihovsky07} from the continuum Feshbach model. The renormalization group analysis of the action \eqref{IRSeff} for nonzero $\kappa$ has been performed in $d=3$ by Frey and Balents \cite{Balents97} at $T=0$, and extended to nonzero temperature by Lee and Lee \cite{Lee04}, revealing a Coleman-Weinberg phenomenon. Thus, for a generic $\kappa \neq 0$ the phase transition will be driven first order. This scenario is realized in the low density limits $n\approx0,2$, where our conclusion thus matches the expectations from the continuum, which was anticipated in \cite{Radzihovsky04,Sachdev04} and discussed in detail in \cite{Radzihovsky07}.

However, the lattice offers the possibility to penetrate the regime where $n\approx 1$. Here, an intriguing situation appears: There exists a point in the phase diagram at which the coefficient of the cubic terms \emph{vanishes exactly}, which happens due to the zero crossing of the coupling $\kappa$. From Eq. (\ref{kappadef}) we have $\kappa \sim c^2 - s^2$. Working with the mean field equation of state $n=2s^2$, one concludes that this takes place at $n=1$. In reality, renormalization effects will add contributions to the naive value of $\kappa$. Furthermore, inspection of the full equation of state \eqref{FullDens} suggests further shifts from the naive expectation, but we have seen in Sec. \ref{sec:ASFDSFex} that close to $n=1$ these are small. Thus, we expect  the decoupling point to be located in the close vicinity of the commensurate point $n=1$. We provide further evidence for this expectation from a symmetry argument in the next section. 


\subsection{Symmetry argument for the Ising quantum critical point} 
\label{sec:SymmArg}

The decoupling of Goldstone and Ising mode at a special point in the phase diagram can also be obtained from a symmetry argument. Being based on a combination of the phase locking symmetry between the degrees of freedom $b_1,b_2$ and a temporally local gauge invariance, it complements the above explicit derivative expansion and sheds more light on the origin of the decoupling of Ising and Goldstone physics. 

For this purpose, let us first discuss the temporally local gauge invariance of the Bose-Hubbard Hamiltonian \cite{SachdevBook} in the presence of an infinite three-body repulsion, which is equivalent to the constrained model under consideration here. This adds a local term to the standard Bose-Hubbard Hamiltonian,
\begin{eqnarray}
H = \lim_{\gamma_3 \to \infty} \big[ H_{BH} + \gamma_3 \sum_i \hat n_i (\hat n_i-1)(\hat n_i-2)\big].
\end{eqnarray}
The temporally local gauge invariance results simply from the fact that the Hamiltonian is not explicitly time dependent (while it is spatially non-local, such that a spatially local gauge invariance does not exist). Consequently, the constrained Bose-Hubbard action must take the form  
\begin{eqnarray}
S_{c} = \int d\tau \sum_i (a_i^\dag( \partial_\tau - \mu )a_i + H[a^\dag,a])
\end{eqnarray}
such that the temporally local gauge invariance is expressed as an invariance under
\begin{eqnarray}
a_i \to \exp \mathrm i \lambda(t) a_i, \quad \mu \to \mu + \mathrm i \partial_\tau \lambda(t) .
\end{eqnarray} 
Since our construction must conserve this property, we also require this invariance for the theory defined with \eqref{HtNew}. On the level of the effective action and in Fourier space, this invariance translates into the Ward identity for the effective action
\begin{eqnarray}\label{WI1}
&&- \frac{\partial }{\partial \mu} \frac{\delta^2 \Gamma}{\delta b^\dag_{1/2}(q) \delta b_{1/2}(q)}\Big|_{b_{1/2}=0; q=0}\\\nonumber
&&\qquad
=  \frac{\partial }{\partial (\mathrm i \omega)} \frac{\delta^2 \Gamma}{\delta b^\dag_{1/2}(q) \delta b_{1/2}(q)}\Big|_{b_{1/2}=0; q=0},
\end{eqnarray}
i.e. the coefficient of the linear time derivative must equal the derivative with respect to the chemical potential. Therefore, in a derivative expansion of the effective action, which is appropriate at low energies, we have: 
\begin{eqnarray}\label{LowEff}
\Gamma\hspace{-0.2cm} &=&\hspace{-0.3cm} \int\hspace{-0.1cm} b_{1}^\dag [z_1 \partial_\tau + y_1 \partial_\tau^2 + m_1^2 + ... ] b_{1} + \ell (b_{1}^{\dag\, 2} + b_{1}^2)  \\\nonumber
&&\hspace{-0.3cm} +b_2^\dag [z_2 \partial_\tau + y_2 \partial_\tau^2 + m_2^2  + ....] b_2   + h (b_{2}^\dag b_{1}^2 + b_{2} b_{1}^{\dag\, 2}) + ...
\end{eqnarray}
The presence of a condensate for $b_2$, $\theta \neq 0$, generates off-diagonal terms in the $b_1$ inverse propagator, i.e. $\ell \neq 0$. Here we restrict to the spatially local part of the effective action, since this is the sector where the coupling of Ising to Goldstone mode emerges. The Ward identity \eqref{WI1} implies $z_{1/2} = -\partial m_{1/2}^2/\partial \mu =: g_{1/2}$.  

Furthermore, using solely the global gauge invariance, we can make the connection between $g_2$ and $g_1$. 
Indeed, we have a phase locking in the $K^{(21)}K^{(10)\,\,\dag} + h.c.$ term. As a consequence of these terms, the phases of $b_1$ and $b_2$ cannot transform independently, and we have 
\begin{eqnarray}
b_{1,i} \to \exp \mathrm i \lambda b_{1,i}, \quad b_{2,i} \to \exp 2 \mathrm i \lambda b_{2,i},
\end{eqnarray} 
leading to the additional Ward identity 
\begin{eqnarray}\label{WI2}
 2 \frac{\partial }{\partial \mu} \frac{\delta^2 \Gamma}{\delta b^\dag_{1}(q) \delta b_{1}(q)}\Big|_{b_{1/2}=0; q=0}\hspace{-0.3cm}
=  \frac{\partial }{\partial \mu} \frac{\delta^2 \Gamma}{\delta b^\dag_{2}(q) \delta b_{2}(q)}\Big|_{b_{1/2}=0; q=0},\nonumber\\
\end{eqnarray}
or $g_2=2g_1$. 
In sum, we have the following relations: 
\begin{eqnarray}\label{SymmRel}
z_2 = g_2 = 2 z_1 = 2 g_1 .
\end{eqnarray}

Next we discuss properties of the ``compressibility'' coupling $g_2(n) = -\partial m_2^2/\partial\mu |_n$, which fixes how strongly the bound state excitation couples to the chemical potential. In the limits $n=0,2$ we can compute it exactly from the solution of the corresponding two-body problems Eqs. \eqref{FullGreen}. At $n= 0$, we find $g_2>0$, while at $n= 2$ we obtain $g_2<0$. These opposite signs can be expected, as at $n=0$ the excitations are well-defined dimers, while at $n=2$ we face well defined di-holes. If we do not redefine the chemical potential, then adding a di-hole is energetically equivalent to delete a dimer. Under the mild assumption that the compressibility is a continuous monotonic function of $n$ (our description is tailored to describe the DSF phase including the phase border, and therein we do not expect additional phase transitions), then $g_2(n)$ must have a unique zero crossing. We note that we should use the above derivative prescription as an operational definition of $g_2$; in principle, there could be a $\mu$-independent constant adding to the full mass or gap term of $b_2$. For $b_1$, such a situation takes actually place and we have an additional mass or gap term $U$. 

As a consequence of Eq. \eqref{SymmRel}, a zero crossing of $g_2$ also implies a zero crossing of the coefficients $z_2,g_1,z_1$. Thus, the leading frequency dependence is not linear, but quadratic, and the analogous statement is valid in the time domain, where the leading behavior is a quadratically appearing time derivative.  

With this result, we now discuss the possible form of the coupling of the Ising to the Goldstone mode. As above, we decompose linearly into massive and phase mode, and absorb the phase fluctuations into dressed $b_1$ fields,  $b_{1} \to \tilde b_{1} = b_{1} (1  - i \tilde \kappa \pi/2^{3/2}), \quad \tilde\kappa = h/\ell$. Indeed the low energy effective action can only depend on derivative couplings associated to the phase mode $\pi$, due to the global $U(1)$ invariance under transformations $\pi \to \pi +\lambda$. The transformaton cancels the cubic term in Eq. \eqref{LowEff} associated to phase fluctuations, while the contribution associated to the real part $\sigma$ can be dropped at low energies since the amplitude is massive. At the same time, the $b_1$ part in the dressed frame now reads 
\begin{eqnarray}\label{LowEff2}
\Gamma_1 &=& \int \tilde b_{1}^\dag [z_1 \partial_\tau + y_1 \partial_\tau^2 + m_1^2 + ... ] \tilde b_{1} \\\nonumber
&&\qquad +\mathrm i z_1\tilde \kappa \partial_\tau\pi \tilde b_{1}^\dag\tilde b_{1}  + \mathrm i y_1\tilde \kappa \partial_\tau^2 \pi \tilde b_{1}^\dag\tilde b_{1}+ ...
\end{eqnarray}
Thus, for $g_2=0$, Eq. \eqref{SymmRel} also implies that the cubic derivative coupling $z_1\tilde\kappa$ with canonical dimension $(3-d)/2$ vanishes. The leading term is a cubic coupling with \emph{quadratic} time derivative. This coupling has canonical dimension $(1-d)/2$, and thus is irrelevant near a Gaussian fixed point for $d>1$. Similarly, a potential $U(1)$ symmetric coupling term $g' \int (\partial_\tau\pi)^2 \phi^2$ has canonical dimension $1-d$. Both therefore do not lead to a  Coleman-Weinberg phenomenon. In consequence, Goldstone and Ising physics effectively decouple at low energies, giving rise to a second order Ising transition. 

We summarize our result. Based on the zero crossing of $g_2$, phase locking and temporally local gauge invariance we find:\\
(i) At the zero crossing point, the nonrelativistic time derivative terms vanish. In the sense of a derivative expansion, the next relevant term is $\partial_\tau^2$, in which case the theory acquires a relativistic space-time isotropy in a $d+1$ dimensional space-time. This is physically sound, as this point has a special kind of (di-)particle-hole symmetry, in that the hybrid excitation consists of  a superposition of ``dimers'' and ``di-holes'' to equal parts. However, we note the absence of a particle-hole symmetry in the conventional sense -- such a situation only occurs in the perturbative limit $J/|U|\to 0$, as discussed in Sec. \ref{sec:DSFCDW}. Beyond the leading order perturbation theory, this symmetry is broken. One manifestation of the absence of this symmetry is the asymmetry of the critical line in the phase diagram, cf. Fig. \ref{PhaseBorder}. \\
(ii) The cubic coupling of Goldstone to Ising mode also vanishes at this point. Only terms which are irrelevant in $d>1$ dimensions then can couple these modes. As a consequence, the Coleman-Weinberg mechanism is suppressed.

We observe that  the constraint influences the physics even at very long wavelengths: It is responsible for the existence of a maximum filling, in turn leading to the existence of a zero crossing of the dimer compressibility, in turn responsible for the existence of the Ising quantum critical point. 

In conclusion, close to the ``particle-hole symmetric'' point at $n=1$, there is a $d+1$ dimensional Ising quantum critical point. Examples of physical realizations of Ising quantum critical points in nature are actually rare. Several systems exhibit Ising type phase transitions with discrete symmetry breaking, like the ASF-DSF transition in the continuum Feshbach model \cite{Radzihovsky04,Sachdev04} and or a transition between superconductors with different pairing symmetries \cite{Vojta00}, but in these cases in the long wavelength limit a Coleman-Weinberg phenomenon takes place. A cubic coupling of the Goldstone mode with \emph{linear} time derivative to the Ising density is actually quite generic in nonrelativisic systems, where the Ising mode emerges as an effective degree of freedom describing the transition from one ordered phase to the other. Here we have identified a mechanism that suppresses this coupling. One of the few other examples for Ising quantum criticality is possibly provided by the model magnet LiHoF$_4$ \cite{Bitko96}, though the issue is debatable due to the long range interactions in the material, preventing an exact mapping to the Ising model. 

The fact that qualitative aspects of the critical behavior are changed in the vicinity of the particle-hole symmetric point $n=1$ bears some resemblance to the physics at the tip of the Mott lobe in the repulsive Bose-Hubbard model. There, the behavior changes from the nonrelativistic $O(2)$ (or XY) universality class with dynamical exponent $z=2$ to the relativistic $O(2)$ model with $z=1$ \cite{FisherFisher89}.

\subsection{Estimate of the Correlation Length}

To get an impression of the perspective to observe Ising quantum criticality in this system experimentally, we estimate the correlation length. This quantity is accessible with current experimental technology \cite{Esslinger07}, and has been measured in continuum Bose gases to characterize critical behavior. 

The Coleman-Weinberg phenomenon manifests itself in the presence of ``runaway'' trajectories on the RG flow diagram. We therefore can estimate the correlation length at the first order phase transition as a scale $l_{\ast }$, at which the runaway trajectory with the corresponding initial conditions hits the boundary of the stability region of the system \cite{AmitBook}. The instability is characterized by the quartic Ising coupling $\lambda$ turning negative, i.e. the condition $\lambda (l _{\ast })=0$.

The scaling properties of the action \eqref{IRSeff} are determined by three parameters:
$V=(\xi/\xi_{+})\sqrt{Z_{\varphi}/Z}$, $U=4!\lambda/\xi_{+}^{3}\sqrt
{Z_{\varphi}}$, and $K=\kappa^{2}\xi^{2}/\xi_{+}^{3}Z\sqrt{Z_{\varphi}}$ with
the corresponding RG equations derived in Ref. \cite{Balents97}. The quantity $V$ scales to zero, therefore we can put $V=0$ from the very beginning. Then the RG equations for the remaining constants $K$ and $U$ read%
\begin{align}
\frac{1}{K}\frac{dK}{dl} &  =\varepsilon-\frac{1}{4}U-\frac{5}{2}%
K,\label{RGoriginal1}\\
\frac{1}{U}\frac{dU}{dl} &  =\varepsilon-\frac{3}{8}U-6K-24\frac{K^{2}}%
{U},\label{RGoriginal2}%
\end{align}
where $\varepsilon=3-d$. To solve these equations, we first introduce new
functions $k  =K\exp(-\varepsilon l), \, u   =U\exp(-\varepsilon l)$, 
and a new variable $x=\exp(\varepsilon l)$. The equations then have the following form,%
\begin{align}
\varepsilon\frac{dk}{dx} &  =-\left(  \frac{1}{4}uk+\frac{5}{2}k^{2}\right)
,\label{RGnew1}\\
\varepsilon\frac{du}{dx} &  =-\left(  \frac{3}{8}u^{2}+6uk+24k^{2}\right)
.\label{RGnew2}%
\end{align}
Writing $u=kf(k)$ and, therefore $du/dk=f+kf^{\prime}$, we obtain%
\begin{equation}
k\frac{df}{dk}=\frac{du/dx}{dk/dx}-f=\frac{f^{2}+28f+192}{2f+20}=\frac{(f+12)(f+16)}{2(f+10)}.\label{eqf}%
\end{equation}
This equation can easily be solved with the result%
\begin{equation}
\frac{k}{k_{0}}=\left(  \frac{f+16}{f_{0}+16}\right)  ^{3}\left(  \frac
{f_{0}+12}{f+12}\right)  ,\label{solutionf}%
\end{equation}
where $f_{0}=u_{0}/k_{0}$ is the initial value for the function $f$ when
$k=k_{0}$.

It follows from Eq. (\ref{RGnew2}) that%
\begin{align*}
\varepsilon\frac{du}{dx}  &  =-\frac{k^{2}}{8}\left[  sf^{2}+48f+192\right] \\
&  =\varepsilon\frac{d}{dx}[kf(k)]=\varepsilon\frac{df}{dx}\left[  f\frac
{dk}{df}+k\right]
\end{align*}
and, after using Eq. (\ref{eqf}), we obtain%
\begin{align*}
\varepsilon\frac{df}{dx}  &  =-\frac{k}{8}(f+16)(f+12)\\
&  =-\frac{1}{8}k_{0}(f_{0}+12)\frac{(f+16)^{4}}{(f_{0}+16)^{3}}.
\end{align*}
The solution of this equation reads%
\begin{equation}
-\frac{x-1}{\varepsilon}\equiv-\frac{\exp(\varepsilon l)-1}{\varepsilon}%
=\frac{8}{3k_{0}}\frac{1}{f_{0}+12}\left[  1-\left(  \frac{f_{0}+16}%
{f+16}\right)  ^{3}\right]  \label{solutionx}%
\end{equation}
and, together with Eq. (\ref{solutionf}), provides a general solution of the
RG equations (\ref{RGnew1}) and (\ref{RGnew2}) and, therefore
(\ref{RGoriginal1}) and (\ref{RGoriginal2}).

The above solution allows us to find the scale $l_{\ast}$, at which the RG
flow reaches the border of stability, $U(l_{\ast})=0$. In $3D$ we obtain
(after taking the limit $\varepsilon=3-d\rightarrow0$)%
\[
-l_{\ast}=\frac{8}{3k_{0}(f_{0}+12)}\left[  1-\left(  1+\frac{f_{0}}%
{16}\right)  ^{3}\right]
\]
with $f_{0}=u_{0}/k_{0}$. As a result, close to the Ising critical point,
$k_{0}\rightarrow0$, we get%
\[
l_{\ast}\sim k_{0}^{-3}\sim(1-n)^{-6}\text{.}%
\]

This result indicates a rather broad critical domain in density around the true Ising critical point, in which the correlation length extends over the whole system. Such extended quasi-critical behavior can be expected for a fluctuation induced first order transition, which results exclusively from the competition of very long wavelength degrees of freedom, and therefore should be weak. For example, already at filling $n=1/4, 2-1/4$ the correlation length is on the order of 15 lattice sites, and greatly exceeds the typical size of an optical lattice of $20$ to $100$ sites at filling $1/2, 3/2$ already by a factor of 10. We conclude that the Ising quantum critical behavior should be experimentally observable in our system. 

Finally, we emphasize that the discussion presented in this section crucially hinges on the fact that our field theoretic setup allows to fully assess the effects of interactions, i.e. nonlinearities in the effective action. Here we have shown that these interaction effects persist even down to arbitrarily long wavelengths. Obviously, such a scenario is not captured in a simple quadratic spin wave theory with \emph{a priori} decoupled atomic and dimer excitations.

\section{Conclusion}
\label{sec:Conclusion}

In this paper, we have performed a detailed analytical investigation of the phase diagram of the attractive lattice Bose gas with a 3-body hardcore constraint. For this purpose, we make use of a method presented in \cite{Diehl09I} which allows to exactly map the constrained model to a theory for two unconstrained bosonic degrees of freedom with conventional polynomial interactions. Within this framework, we particularly focus on effects tied to interactions, which cannot be addressed within a mean field plus spin wave approach. While our analysis confirms the rough features of the phase diagram obtained from a simple mean field approach -- the presence of an Ising-type phase transition from an atomic to a dimer superfluid, numerous interaction driven effects are identified. These arise on various length scales, ranging from the fluctuation induced formation of the dimer (or di-hole) bound state on top of the vacua at $n=0$ and $n=2$ on the microscopic level over a an understanding of the beyond mean field effects causing nonuniversal shifts in the phase boundary and giving rise to the proximity of the system to a bicritical point with enhanced $SO(3)$ symmetry in strong coupling, down to the assessment of the true nature of the phase transition at very long wavelength. This underpins the fact that short and long range correlations can then be treated within a unified formalism. 

\emph{Acknowledgements} -- We thank E. Altman, A. Auerbach, H. P. B\"uchler,  M. Fleischhauer, M. Greiter, A. Muramatsu, N. Lindner,  J. M. Pawlowski, L. Radzihovsky, S. Sachdev, J. Taylor and C. Wetterich for interesting discussions. This work was supported by the Austrian Science Foundation (FWF) through SFB F40 FOQUS, and project I118\_N16 (EuroQUAM\_DQS), by the European union via the  integrated project SCALA, by the Austrian Ministry of Science BMWF via the UniInfrastrukturprogramm of the Forschungsplattform Scientific Computing and of the Centre for Quantum Physics, by the Russian Foundation for Basic Research, and by the Army Research Office with funding from the DARPA OLE program.

\end{document}